\documentclass[11pt]{article}
\pdfoutput=1

\usepackage[utf8]{inputenc}
\usepackage[margin=1in]{geometry}
\usepackage{graphicx}
\usepackage{booktabs}
\usepackage{multirow}
\usepackage{amsmath,amssymb,amsfonts}
\usepackage{mathrsfs}
\usepackage[title]{appendix}
\usepackage{xcolor}
\usepackage{textcomp}
\usepackage{manyfoot}
\usepackage{algorithm}
\usepackage{algorithmicx}
\usepackage{algpseudocode}
\usepackage{listings}
\usepackage[detect-all]{siunitx}
\usepackage{makecell}
\usepackage{xspace}
\usepackage{array}
\usepackage{natbib}

\PassOptionsToPackage{hyphens}{url}
\usepackage[colorlinks,allcolors=black]{hyperref}

\newcolumntype{L}[1]{>{\raggedright\arraybackslash}p{#1}}

\newcolumntype{P}[1]{>{\raggedright\arraybackslash}p{#1}}

\newcommand{\name}{DRAGON\xspace}

\def\fIScoreUs{60.8\%\xspace}
\def\fIScoreLegion{54.8\%\xspace}
\def\datasetRepos{\num{825 561}\xspace}

\begin{document}

\title{\name: Robust Classification for Very Large Collections of Software Repositories}

\author{
  Stefano Balla$^{1}$ \and
  Stefano Zacchiroli$^{2}$ \and
  Thomas Degueule$^{3}$ \and
  Jean-Rémy Falleri$^{3,4}$ \and
  Romain Robbes$^{3}$\\
  \\
  $^{1}$DISI – Università di Bologna, Italy\\
  $^{2}$LTCI, Télécom Paris, Institut Polytechnique de Paris, Palaiseau, France\\
  $^{3}$Univ. Bordeaux, CNRS, Bordeaux INP, LaBRI, UMR 5800, France\\
  $^{4}$Institut Universitaire de France, France\\
  \\
  \texttt{stefano.balla2@unibo.it, stefano.zacchiroli@telecom-paris.fr,}\\
  \texttt{thomas.degueule@labri.fr, falleri@labri.fr, romain.robbes@labri.fr}
}

\date{}

\maketitle

\begin{abstract}

The ability to automatically classify source code repositories with ``topics'' that reflect their content and purpose is very useful, especially when navigating or searching through large software collections.
However, existing approaches often rely heavily on README files and other metadata, which are frequently missing, limiting their applicability in real-world large-scale settings.

We present \name, a repository classifier designed for very large and diverse software collections. It operates entirely on lightweight signals commonly stored in version control systems: file and directory names, and optionally the README when available.
In repository classification at scale, \name improves F1@5 from \fIScoreLegion to \fIScoreUs, surpassing the state of the art.

\name remains effective even when README files are absent, with performance degrading by only 6\% w.r.t.~when they are present.
This robustness makes it practical for real-world settings where documentation is sparse or inconsistent.
Furthermore, many of the remaining classification errors are near misses, where predicted labels are semantically close to the correct topics.
This property increases the practical value of the predictions in real-world software collections, where suggesting a few related topics can still guide search and discovery.

As a byproduct of developing \name, we also release the largest open dataset to date for repository classification, consisting of 825 thousand repositories with associated ground-truth topics, sourced from the Software Heritage archive, providing a foundation for future large-scale and language-agnostic research on software repository understanding.

\end{abstract}


\section{Introduction}
\label{sec:intro}

Exhaustive collections of software source code repositories have grown to hundreds of millions. GitHub alone reports more than 420 million repositories, both public and private, in early 2025.\footnote{\url{https://github.com/about}, accessed 2025-03-10 (like other URLs in the paper, unless otherwise stated)} Software Heritage, abbreviated as SWH \citep{ipres-2017-software-heritage}, the largest public archive of software source code, has archived more than 350 million.\footnote{\url{https://archive.softwareheritage.org}} The World of Code, known as WoC \citep{ma2019woc}, is in a similar ballpark.
Browsing, searching through, and discovering relevant projects in such vast collections remain very challenging to this date.
This limitation impacts both professional software developers looking for open source components to reuse, and empirical software engineering researchers assembling large sets of repositories to analyze.

\emph{Repository classification} \citep{secil_whats_code2002,kawaguchi_mudablue_2004} associates descriptive labels (or ``topics'') with software repositories to support navigation and discovery.
Most existing approaches use the topics that GitHub users assign to their own repositories\footnote{\url{https://github.com/topics}} as ground truth to train \emph{automated} classifiers, which can then label repositories that lack such metadata \citep{di_rocco_topfilter_2020,di_rocco_hybridrec_2023,dang_legion_2024,izadi_topic_2021,izadi_semantically-enhanced_2023}.
However, these user-provided topics cover only a small portion of the ecosystem: roughly 4\% of repositories on GitHub and just 2\% of those archived in Software Heritage include them.
This scarcity underscores the need for automated classifiers that can assign meaningful topics to the vast majority of unlabeled repositories, enabling large-scale search and analysis.

However, State-of-the-art repository classification methods perform poorly in very large and heterogeneous collections such as SWH and WoC. Three main limitations stand out (discussed further in Section~\ref{sec:motivation}):

\begin{enumerate}
    \item \textbf{Topic quality.} GitHub topics are user-generated and often noisy \citep{sas_antipatterns_2022}. They frequently include programming languages, which are trivial to infer via file extensions \citep{fratantonio2024magika} and inflate reported accuracy. Focusing on high-value applications and domain topics (e.g., \emph{game}, \emph{database}, \emph{computer configuration}) is both harder and more useful, but is rarely studied at large scale \citep{leclair_adapting_2018,ma_automatic_2018,yusof_classification_2010,soll_classifyhub_2017,linares-vasquez_using_2014,mcmillan_categorizing_2011}.

    \item \textbf{Platform dependency.} Several classification approaches rely on metadata such as repository descriptions and wiki pages, which are not recorded in version control and thus absent from repository collections like SWH and WoC. This limits their applicability beyond GitHub.

    \item \textbf{README dependence.} Many classification methods rely heavily on README content, yet README files are missing in 42\% of repositories in SWH—over 140M projects. A classifier whose performances degrade sharply without README is ill-suited for large-scale archives.
\end{enumerate}

To address these challenges, we introduce \textbf{\name}, a new \emph{robust and scalable multi-label repository classifier} that tags repositories with \emph{high-value topics} from the curated GitRanking taxonomy \citep{sas_gitranking_2023}.

\name is trained on a novel open dataset of over \num{825 000} repositories extracted from Software Heritage and 239 topics from GitRanking (excluding programming languages). The dataset, presented in Section \ref{sec:dataset}, is the largest ever used for repository classification, 5.5$\times$ larger than previous ones.

\name integrates two sources of information: (1) file and directory names from the source tree and (2) the README file, if available. It employs a \emph{sentence-pair BERT} model \citep{devlin_bert_2019}, which keeps file tree and textual signals distinct.
A key design objective of \name is robustness, understood as the ability to preserve most of its predictive performance when README content is missing.
This property is essential because real-world repositories vary widely in documentation quality, and many entirely lack a README. At the same time, achieving robustness without retraining ensures that a single model can operate seamlessly across both README-rich and README-scarce cases, avoiding the need to maintain separate models for different input conditions. This combination makes \name more practical and directly applicable to large, real-world repository collections.

The dataset used for the training is highly imbalanced, with a few very frequent topics such as \emph{computer configuration} or \emph{game} and many rare ones. This long-tailed distribution makes learning difficult for infrequent classes. We mitigate this class imbalance using \emph{focal loss} \citep{Lin_focal_loss2017}, emphasizing harder examples during training, and refine multi-label outputs by comparing global \citep{lipton_thresholds_2014} and per-class thresholding strategies \citep{Narasimhan_class_cov_2021}. The model architecture and training procedure are described in detail in Section \ref{sec:methodology}; the experimental setup is reported in Section \ref{sec:experimental_setup}.

\subsection{Research Questions}
To evaluate \name, we investigate the following research questions:

\begin{itemize}
    \item \textbf{RQ1 (Effectiveness on domain topics):} How accurately can repositories be classified on domain-only labels (excluding those related to languages) at scale?\\

    \item \textbf{RQ2 (Robustness and design under README scarcity):} How robust is performance when README is absent at inference \emph{without retraining}, and does sentence-pair encoding outperform single-sequence concatenation when README is missing?\\

    \item \textbf{RQ3 (Evaluation protocol):} Which thresholding strategy best balances performance and class coverage?\\

    \item \textbf{RQ4 (Error characteristics):} Which kinds of errors occur, and how much are they due to semantic near-misses?\\
\end{itemize}

\subsection{Results}
\name achieves F1@5 score of \fIScoreUs, surpassing the best previous approach in the literature \citep{dang_legion_2024} by 6\%.
When README files are missing, a common occurrence in our target use cases, \name, \emph{without retraining}, degrades by only 6\%, versus -9.5\% in the state-of-the-art with retraining \citep{izadi_topic_2021}.

Beyond aggregate metrics, our error analysis shows that a large portion of apparent misclassifications are semantically close to the correct topics—so-called near misses. This makes the predictions useful in practice even when exact matches differ from the ground truth, since they still reflect the project’s domain with meaningful proximity.

Sections~\ref{sec:results} and \ref{subsec:error_similarity} details these findings, which we discuss in Section~\ref{sec:class_discussion} before concluding in Section~\ref{sec:conclusion}.

\subsection{Data availability}
A complete replication package \citep{replication-package} for this work is available from Zenodo.

\section{Background and Motivation}
\label{sec:motivation}

We articulate the constraints of exhaustive archives (resource limits, forge heterogeneity, and missing documentation), motivating our design choices in Section \ref{sec:methodology} and the evaluation criteria in Section \ref{sec:experimental_setup}.

Large-scale project classification has several valuable use cases. It allows to refine navigation and searches for relevant projects in large-scale software repositories and ecosystems, which is useful for developers searching for projects as well as scientists studying the practice of software engineering. Applied to exhaustive software archives such as Software Heritage, it allows for more representative empirical studies, which is important for generalizing findings~\citep{nagappan2013diversity}. \citet{trujillo2022penumbra} found that open-source projects \emph{not} on GitHub differ from the open-source projects found on GitHub: notably, projects not on GitHub tended to be maintained for longer, had more collaborators, and focused on different domains. Such exhaustiveness further allows for larger-scale studies of specific categories of software repositories.

\subsection{Assumptions on data availability}
While several project classification approach exist, they take assumptions that are not suitable for such large-scale scenarios targeting an exhaustive software archive.

In truly large scale scenarios, accessing all the content is an expensive operation, due to the sheer size of it.
For instance, a copy of Software Heritage, the largest public source code archive, currently requires about 2 PiB of storage.
Just storing this amount of data was estimated to cost around 50,000 US\$ in 2023.\footnote{\url{https://ourworldindata.org/grapher/historical-cost-of-computer-memory-and-storage}}

This provides a strong incentive to minimize the resources required to build an automatic repository classification system.
Classification approaches that require access to source code cannot scale to such volumes of data.

Not all software forges have the same metadata; approaches that work on GitHub may rely on metadata specific to GitHub, such as project descriptions of wiki pages, that are not available in other forges.
Topics, which may be used to train a project classifier, are not present in all the forges.

Taking into account the diversity of metadata and the size constraints, a common set of features naturally emerges: the file tree data of the project (in terms of its name and its file names) as a lightweight approximation of its content; as well as the README file, a high-signal source of high-level documentation about the project.
Downloading and storing a single and small text file is a much more scalable solution in a large-scale scenario.

Even relying on this common denominator is not valid in many cases.
In Software Heritage, 42\% of projects lack a README file.
This could be for several reasons.
The main documentation file (the README) may be found in different locations in different forges, and writing heuristics to find it has diminishing returns.
The project may use other forms of documentation that are hard to detect via heuristics, that, by necessity, can not look at the content of entire project.
Even when a README is found, it can be difficult to use for multiple reasons: it can be very terse (e.g., if other sources of documentation are available); a significant portion of projects may be written in languages other than English, or that use encodings other than ASCII.
Several project classification approaches filter out examples whose README do not contain enough english data, e.g. \citep{dang_legion_2024,izadi_semantically-enhanced_2023}, once again limiting the applicability of these approach.
Thus, an approach that works on exhaustive archives should have good results even in the absence of a README.
While previous works illustrate the feasibility of multi-label classification using textual metadata, existing approaches suffer from severe performance degradation \citep{izadi_topic_2021} or become entirely inapplicable \citep{izadi_semantically-enhanced_2023,widyasari_topic_2023} when README files are absent.
The project file tree (directory and file names) is always present and retrievable, while being also a small source of information.

\subsection{Assumptions on label quality}

In addition, many project classification approaches make assumptions about label quality.

A convenient source of labels for projects are GitHub Topics and GitHub Featured Topics.
GitHub Topics \footnote{\url{https://docs.github.com/en/repositories/managing-your-repositorys-settings-and-features/customizing-your-repository/classifying-your-repository-with-topics}} are user-defined tags that repository maintainers assign to describe the content, language, or domain of a project (e.g., machine-learning, bioinformatics) .
Featured Topics \footnote{\url{https://github.com/topics/featured}} are a curated subset of these topics, selected and maintained by GitHub staff.
However, this data source features a large amount of variability in terms of quality of the data, distribution of the data among classes, and in terms of the domains of the labels.
Most recent approaches rely on Github Featured Topics, which feature a level of data curation, as opposed to regular GitHub topics.
However, Featured Topics are still low quality classes (e.g., terse topics such as \#rna-seq, or \#ci), as demonstrated by \citet{sas_gitranking_2023} and \citet{sas_antipatterns_2022} who explored their inconsistencies and limitations.

In particular, a large portion of such labels represent programming languages (e.g., a web library written in python may have the topics \#web and \#python).
Classifying programming languages in this way impacts the performance of the classifier in several ways: first, these labels might be often very easy classes to guess (especially if a classifier can access file names) and they are also common (see Table \ref{tab:topics-comparison}).
This leads to an over-optimistic performance on datasets that feature these programming language topics.
Second, the model dedicating some of its capacity to these labels may lead it to allocate less capacity for other labels (less common and more difficult), negatively affecting performance there.
Finally, the task of predicting programming languages is not necessarily a machine learning task, as many tools exist for this very topic, from the elaborate such as Guesslang \footnote{\url{https://github.com/yoeo/guesslang}} and Linguist \footnote{\url{https://github.com/github-linguist/linguist}}), to more crude heuristics based on file extensions.

More generally, performance on a few labels should not dominate the overall performance, and should not be at the detriment of the performance of the other labels.
A project classifier that targets exhaustive software archives should have performance that is well distributed among all possible classes.

To sum up, the choice of labels in the classification has a major impact on the actual task that the classifier solves (i.e., whether it is a practical task or not) and on the performance of the classifier (and whether it reflects reality).
Therefore, the selection of labels should be approached with methodological rigor and clear justification.

\citet{sas_gitranking_2023} introduced an active-sampling approach to identify and create a taxonomy of the most informative topics for repository labeling.
This process condensed 130k Github topics into a well-structured taxonomy of 301 topics, specifically focusing on application domains while excluding programming languages, a trivial category to predict, as previously demonstrated by \citet{di_sipio_multinomial_2020}.
The resulting taxonomy, named GitRanking, provides a structured mapping from raw GitHub topics to curated taxonomy topics, allowing for the classification of repositories based on more meaningful topic representations.
The authors also determine the depth of each topic within the taxonomy, specifying the hierarchical level at which a topic resides.
This taxonomy should serve as a foundational reference for any repository classification problem.
Following this taxonomy creation, \citet{sas_gitranking_2023} applied a probabilistic methodology, inspired by their prior work \citep{di_sipio_multinomial_2020}, to solve the multi-class classification problem.
While Sas et al. made an important first step toward improving label quality, their methodology did not yet achieve state-of-the-art performance: they report an F1 score of 34\%, which is below the results achieved by later approaches discussed in Section~\ref{sec:related-work} and summarized in Table~\ref{tab:sota_all}.
It is worth noting, however, that \citet{sas_gitranking_2023} are the only prior work that, like our study, removes programming language labels from the taxonomy. This makes their classification task substantially harder and renders direct comparison with models evaluated on language-heavy topic sets less meaningful.

\bigskip
Building on this foundation, our work addresses the remaining limitations by targeting robustness under sparse documentation and scalability to exhaustive archives—challenges that prior approaches, even with curated taxonomies, have yet to resolve.
The following section situates our contribution within this broader landscape of repository classification methods.

\section{Related work}
\label{sec:related-work}

Guided by \textbf{RQ1–RQ4}, we analyze prior work along three axes:
\emph{(i)} availability of input features across platforms and archives,
\emph{(ii)} quality and nature of the label taxonomy, and
\emph{(iii)} scalability to collections containing hundreds of millions of repositories.
Within this space, two main paradigms have emerged:
\emph{(A) source-code-based methods} and
\emph{(B) textual or metadata-based methods}.
These paradigms differ in the signals they exploit (code vs.\ documentation), the type of learning (unsupervised vs.\ supervised), and the scale of datasets they can handle.

\subsection{Source-code-based methods}
Many early studies attempted to infer domain or functionality from raw source code.
For instance, Kawaguchi et al.\ introduced MUDABlue \citep{kawaguchi_mudablue_2004}, applying Latent Semantic Analysis (LSA) to categorize SourceForge repositories of only a few thousand projects.
Others employed latent Dirichlet allocation (LDA) on code bases \citep{linstead_mining_2007,tian_using_2009}, or techniques like SVM and decision trees \citep{linares-vasquez_using_2014,yusof_classification_2010}.
More modern approaches, such as LeClair et al. \citep{leclair_adapting_2018}, leverage Neural Networks on roughly 9k projects with curated domain labels.
While code-level analysis can yield meaningful categories, it is often \emph{expensive to mine} at scale, and so not a feasible solution for our use case.
Most code-centric efforts remain limited to only a few hundred \citep{kawaguchi_mudablue_2004,tian_using_2009} or, at best, a few tens of thousands of repositories \citep{linares-vasquez_using_2014}, making them unsuitable for massive archives like SWH or WoC.

\subsection{Textual and metadata-based methods}
Another extensive line of research leverages \emph{textual documentation} to categorize software.
In supervised multi-label settings, researchers often adopt a \emph{TF-IDF} or embedding-based transformation of textual data and then map it to preexisting topics.
For instance, \citet{izadi_semantically-enhanced_2023} proposed a \emph{semantically-enhanced} recommendation pipeline using Knowledge Graphs and Logistic Regression (LR), trained on approximately 150k GitHub repositories.
\citet{izadi_topic_2021} used DistilBERT or TF-IDF + LR on README, wiki, and file names for multi-label classification, showing that TF-IDF + LR achieves stronger results in practice.
Both studies, however, rely on topic sets that include a substantial proportion of programming languages: in \citet{izadi_semantically-enhanced_2023}, six of the twenty-five most frequent topics are languages, and in \citet{izadi_topic_2021}, nine of the top twenty-five fall in that category.
This inflates apparent performance, since language prediction is a comparatively trivial task.
Although these approaches scale to roughly 150k repositories, they still depend on the presence of documentation (e.g., README files).
In their ablation study, removing the README led to a sharp performance drop, with F1@5 declining from 47\% to 37.5\%.
\citet{widyasari_topic_2023} reformulated the same task, under the same assumptions of complete documentation and language-heavy topic sets, as an extreme multi-label classification problem evaluated on about 22k repositories, reporting improved metrics over previous techniques.
\citet{dang_legion_2024}, building on the same dataset, further approached the task as a long-tail problem, introducing a distributed balance loss and quality filtering mechanism to improve performance metrics.
\citet{sas_gitranking_2023} proposed a smaller but higher-quality dataset of 48k repositories labeled using the curated GitRanking taxonomy, focusing on domain topics rather than programming languages.

\subsection{Positioning and contributions}
The successes of prior work are often tied to two favorable conditions:
\emph{(i)} the use of programming language topics, which are easy to predict and inflate reported performance, and
\emph{(ii)} evaluation on repositories with rich textual documentation, particularly README files.
In contrast, large-scale archives such as SWH contain millions of repositories without README content, and practical discovery tasks require domain-oriented topics rather than programming languages.
This gap motivates our focus on robust classification under sparse textual conditions and with higher-value labels.

{\sloppy
\begin{table*}[t]
\scriptsize
\setlength{\tabcolsep}{3pt}
\renewcommand{\arraystretch}{1.7}
\caption{Comparison of repository classification studies based on dataset characteristics.}
\centering
\begin{tabular}{L{3cm}|l|p{2.8cm}|P{2.3cm}|c}
\toprule
\textbf{Study} & \textbf{Size} & \textbf{Data} & \textbf{Label source} & \textbf{\makecell{Prog. langs.\\as topics}} \\
\midrule
\cite{kawaguchi_mudablue_2004,linstead_mining_2007,tian_using_2009,linares-vasquez_using_2014,yusof_classification_2010} & $<3300$ & Source Code & SourceForge Domains & No \\
\hline
\cite{leclair_adapting_2018} & \num{9000} & Source Code, Project descriptions & Curated Domains & No \\
\hline
\cite{izadi_semantically-enhanced_2023,izadi_topic_2021} & \num{150000} & Readme, Wiki, Project Description, file names & Featured Topics & Yes \\
\hline
\cite{widyasari_topic_2023} & \num{22000} & Readme, Project Description & Featured Topics & Yes \\
\hline
\cite{dang_legion_2024} & \num{22000} & Readme & Featured Topics & Yes \\
\hline
\cite{sas_gitranking_2023} & \num{48000} & Readme, Project Description & GitRanking Topics & No \\
\hline
\emph{Our work} & \num{825000} & File/Repository names, Readme (optional) & GitRanking Topics & No \\
\bottomrule
\end{tabular}
\label{tab:study_comparison}
\end{table*}
}

\bigskip
Table~\ref{tab:study_comparison} summarizes representative repository classification studies by dataset size, input signals, and label sources.
The comparison highlights four persistent issues across prior work:
\begin{enumerate}
\item \emph{Dataset sizes} rarely exceed 150k repositories, and many studies cover only a few thousand.
\item \emph{Overreliance on code or textual documentation} hampers performance when README or other textual fields are incomplete or absent, or make approaches reliant on code inapplicable at a large scale.
\item \emph{Low-quality classes} often yield only superficial or uninformative labels.
\item \emph{Prediction of programming languages} dominates many efforts, even though simpler standalone tools (e.g., Guesslang) can solve that problem more efficiently.
\end{enumerate}

To address these shortcomings, we propose:
\begin{itemize}
\item \emph{The creation and the use of a large-scale, real-world dataset} of over 825k repositories from Software Heritage, surpassing previous studies by a wide margin.
\item \emph{Excluding programming language detection} and focusing instead on a high-quality taxonomy of domain-specific classes (GitRanking).
\item \emph{Combining file tree and textual data} within a sentence-pair BERT architecture, achieving strong predictive power even for repositories that lack a README.
\end{itemize}
This work thus advances the state of the art in both \emph{scale} and \emph{robustness}, delivering a more resilient and domain-focused classification pipeline aligned with the realities of massive open-source ecosystems.

\section{Dataset}
\label{sec:dataset}

\begin{figure}
    \centering
    \includegraphics[width=0.99\linewidth]{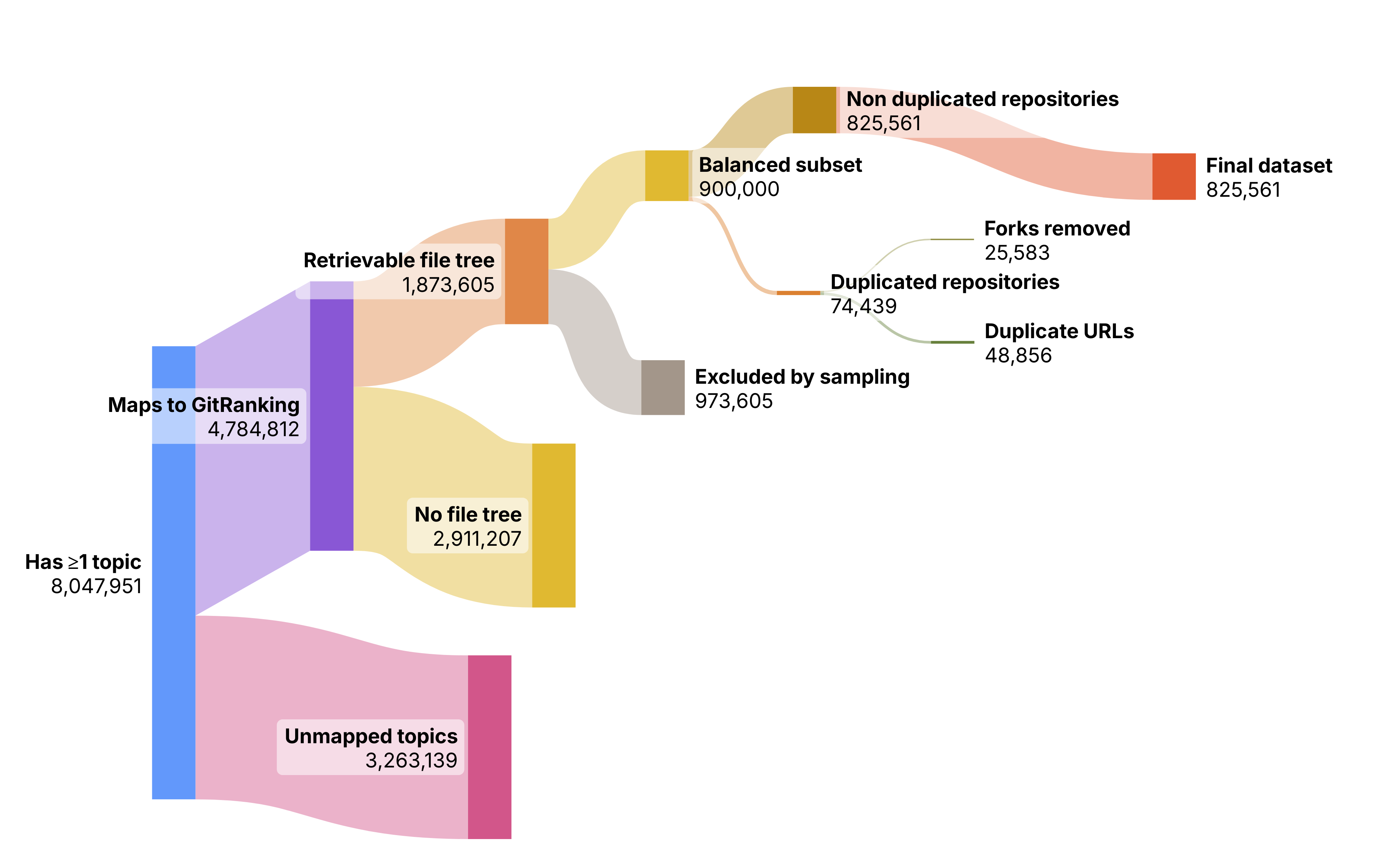}
    \caption{Overview of the data processing and filtering pipeline.
    The Sankey diagram shows the progressive reduction of repositories
    from the initial \num{8047951} GitHub projects archived by Software Heritage with at least 1 topic,
    to the final dataset of approximately \num{825 000} repositories.
    Each split corresponds to a major operation described in the following subsections:
    \ref{topic-filtering} Topic filtering, \ref{repository-mining} Repository mining and \ref{subsec:data_preparation} Data cleaning.}
    \label{fig:sankey_pipeline}
\end{figure}

We detail in this section how we assembled our dataset of software repositories and associated topics, the largest available to date for repository classification tasks.
The main stages of the dataset extraction and cleaning pipeline are summarized in Figure~\ref{fig:sankey_pipeline}.

The dataset design reflects our research questions. It prioritizes domain-oriented topics (RQ1), includes repositories even when README files are absent (RQ2), maintains a long-tail topic distribution to study thresholding trade-offs (RQ3), and preserves sufficient provenance to support a semantic error analysis (RQ4).

\label{subsec:data_extraction}

\subsection{Initial dataset}
We obtained from the Software Heritage archive \citep{ipres-2017-software-heritage} an export of all archived Git repositories hosted on GitHub and associated topics on that platform.
The choice of starting from GitHub repositories is due to the large availability of topics on that platform, but the obtained classifier is independent on any specific platform and only need information stored in version control systems (VCSs) to operate.
The initial dataset contained \num{216 775 237} repository addresses (called ``origins'' in SWH jargon) and the topics associated with each of them.
We also obtained the most recent version of the SWH graph dataset \citep{pietri2019graphdataset}, to be used later for repository mining purposes.

\subsection{Topic filtering}
\label{topic-filtering}
To prepare an annotated dataset for training, we only kept repositories with \emph{at least one associated topic}.
This reduced the dataset to \num{8 047 951} repositories.
Approximately 96\% of GitHub repositories (98\% of all repositories) archived in SWH lack associated topics, strengthening the need to label them in order to peruse this huge collection.

\begin{table}
    \centering
    \caption{Top-30 topics before/after GitRanking mapping, with programming language topics highlighted in \textbf{bold}.}
    \label{tab:topics-comparison}
    \begin{tabular}{@{}l r | l r@{}}
        \toprule
        \multicolumn{2}{c}{\textbf{Before applying GitRanking}} & \multicolumn{2}{c}{\textbf{After applying GitRanking}} \\
        \cmidrule(lr){1-2} \cmidrule(lr){3-4}
        \textbf{Topic} & \textbf{Frequency} & \textbf{Topic} & \textbf{Frequency} \\
        \midrule
        \textbf{javascript} & \num{93384} & computer conf. & \num{93494} \\
        \textbf{r} & \num{83627} & game & \num{49613} \\
        \textbf{python} & \num{81674} & website & \num{42235} \\
        \textbf{c} & \num{75459} & algorithm & \num{29360} \\
        react & \num{62258} & database & \num{22110} \\
        nodejs & \num{48671} & internet bot & \num{18227} \\
        \textbf{java} & \num{44725} & blockchain & \num{17511} \\
        \textbf{css} & \num{44298} & web application & \num{17494} \\
        \textbf{html} & \num{40952} & data structure & \num{15683} \\
        \textbf{typescript} & \num{32724} & video game dev. & \num{15623} \\
        \textbf{d} & \num{29195} & artificial neural net. & \num{14901} \\
        \textbf{php} & \num{27034} & front end & \num{13741} \\
        hacktoberfest & \num{26040} & automation & \num{13167} \\
        android & \num{23411} & security & \num{12927} \\
        docker & \num{23064} & server & \num{12523} \\
        machine-learning & \num{21632} & visualization & \num{12518} \\
        api & \num{18572} & microservices & \num{12163} \\
        \textbf{v} & \num{17778} & software testing & \num{11857} \\
        mongodb & \num{17262} & digital image proc. & \num{11655} \\
        express & \num{14754} & deep neural network & \num{11507} \\
        \textbf{csharp} & \num{13307} & cryptography & \num{10578} \\
        git & \num{13113} & animation & \num{10282} \\
        bootstrap & \num{12913} & cryptocurrency & \num{9977} \\
        vue & \num{12658} & websocket & \num{9976} \\
        redux & \num{12422} & music & \num{9865} \\
        django & \num{12373} & data & \num{9633} \\
        \textbf{cpp} & \num{12338} & mathematics & \num{9479} \\
        mysql & \num{11854} & simulation & \num{8742} \\
        deep-learning & \num{11805} & robotics & \num{8717} \\
        angular & \num{11797} & back end & \num{8675} \\
        \bottomrule
    \end{tabular}
\end{table}

As user-provided topics (those in the initial dataset) tend to be noisy and low-quality \citep{sas_gitranking_2023,sas_antipatterns_2022}, we remapped them to higher-quality ones using the GitRanking taxonomy \citep{sas_gitranking_2023} and associated mapping.
This step also filters out implicitly programming language topics, as they are not part of the GitRanking taxonomy.

Table \ref{tab:topics-comparison} compares the top 30 topics before and after GitRanking mapping, illustrating the prevalence of programming languages (before) and that of higher-value domain topics (after).
Before the mapping, the most frequent topics were dominated by specific technologies and frameworks (e.g., \textit{React, Node.js, Android, Docker, MongoDB, Express, Vue, Django}), reflecting transient or tool-oriented interests typical of user tagging behavior on GitHub.
In contrast, after applying the GitRanking-official remapping, the topic distribution becomes more balanced and oriented toward broader, longer-lived software domains (e.g., \textit{database}, \textit{security}, \textit{blockchain}, \textit{data visualization}).
This shift highlights how GitRanking improves label consistency and conceptual abstraction, better supporting the study of repository functionality rather than popularity of individual technologies.
Nonetheless, technology-specific topics remain valuable indicators of emerging trends and could be investigated in future work to analyze temporal dynamics or framework adoption patterns.
After the mapping we obtain \num{4 784 812} repositories with at least one (GitRanking) topic and a total of 239 distinct topics used across all remaining repositories.
We discard all repositories with no remaining topics.
In summary, this re-labeling directly supports RQ1 by focusing evaluation on good-quality domain topics instead of programming languages.

\subsection{Repository mining}
\label{repository-mining}
We collected three kinds of information from each remaining repository: \emph{file tree data}, \emph{textual data}, and \emph{metadata}:

\begin{itemize}
    \item \textit{File tree data:} to capture the organization of each repository, we extracted its file and directory hierarchy from the HEAD commit (i.e., the most recent commit in each repository).
    We recorded the name (including extensions) of every directory and every file in the HEAD commit, preserving the hierarchical file-system structure.
    We also preserved symbolic links, without following them, to preserve file structure as shown to developers.

    \item \textit{Textual data:} we looked for the presence of README files located at the root directory of the HEAD commit.
    We considered common variants of the \texttt{README} name (e.g., \texttt{README.md}, \texttt{README.txt}, \texttt{README.rst}, ignoring case differences) and excluded files larger than 100\,MiB.
    Each identified README was retrieved from the SWH archive and stored in the dataset based on its intrinsic SHA1 identifier.

    \item \textit{Repository metadata:} finally, we retained from the initial dataset the following repository metadata, as originally stored on GitHub and archived by SWH: URL (e.g., \texttt{https://github.com/user/repo}), whether it is a fork or not (only for deduplication purposes later on), the topics.

\end{itemize}
This repository mining step has been implemented using the compressed graph representation of the SWH archive \citep{boldi2020swhgraph}, via its Rust API.
Code for this step is provided as part of the replication package for this paper \citep{replication-package}.

\subsection{Data cleaning}
\label{subsec:data_preparation}

Some repositories did not have a retrievable file tree in their HEAD commit.
The most common reason for this is that they were empty repositories at the time of archival by SWH.
Excluding them led to \num{1 873 605} repositories with an associated file hierarchy.
Most of the excluded repositories had \texttt{github-config} as their only topic, indicating a repository used to configure GitHub-specific features (e.g., user profiles, wiki pages.

\begin{figure}
    \centering
    \includegraphics[width=0.99\linewidth]{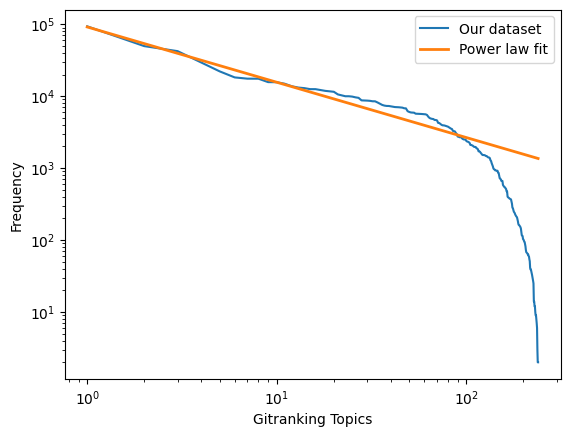}
    \caption{Topic distribution vs. power law ($y \propto x^{-0.77}$).}
    \label{fig:long_tail}
\end{figure}

To address class imbalance, we employed an inverse-frequency sampling approach, where weights were assigned inversely proportional to the frequency of topic combinations. This ensured that less common topics received higher weights in the sampling process.
This approach targeted \num{900 000} repositories, determined iteratively using the elbow method to balance coverage. We adjusted the dataset size until the most frequent topic was not disproportionately dominant over the second most frequent, ensuring a more even representation.
After applying balancing strategies, the final topic distribution, shown in Figure \ref{fig:long_tail}, still exhibits a long-tail pattern closely following a power law, which is desirable as it aligns with real-world data conditions.

To avoid contamination between training and test sets, we deduplicated repositories in the dataset using two approaches.
First, we removed repositories declared as forks on GitHub, which resulted in the removal of \num{25 583} repositories.
Second, we removed repositories having the same URLs, which could be present in the dataset due to SWH archiving the same repositories via different protocols or through renames over time; this resulted in removing a further \num{48 856} repositories.
To break ties when deduplicating, we always kept the most recently archived repository.
After deduplication, we obtain a final dataset of \datasetRepos repositories.

Following \citet{izadi_topic_2021}, we apply a standardized cleaning pipeline also to the content of README files, to enhance consistency and reduce noise.
We remove emails, URLs, usernames, embedded code snippets, punctuation, digits, and non-ASCII characters.
The latter step is necessary as the \texttt{bert-base} tokenizer, which we use, has an English-centric vocabulary and replaces many non-ASCII symbols with \texttt{[UNK]} tokens, leading to data contamination.
Additionally, we strip HTML/XML tags, markdown symbols, and overlong whitespaces.

\begin{figure}
    \centering
    \includegraphics[width=0.99\linewidth]{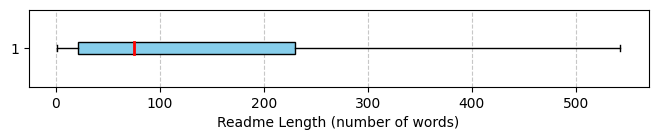}
    \caption{Distribution of README lengths (in words) for our dataset, excluding outliers. Most repositories have minimal documentation.}
    \label{fig:readme}
\end{figure}

\subsection{Dataset statistics}
\label{par:stats}
The final dataset contains \datasetRepos repositories, with \num{277 907} (34\%) of them lacking a README file.

READMEs in the dataset tend to be short, with a median length of 75 words (see distribution in Figure \ref{fig:readme}): 25\% are under 20 words, and 75\% contain fewer than 200, with very few exceeding 500.

\begin{figure}
    \centering
    \includegraphics[width=0.99\linewidth]{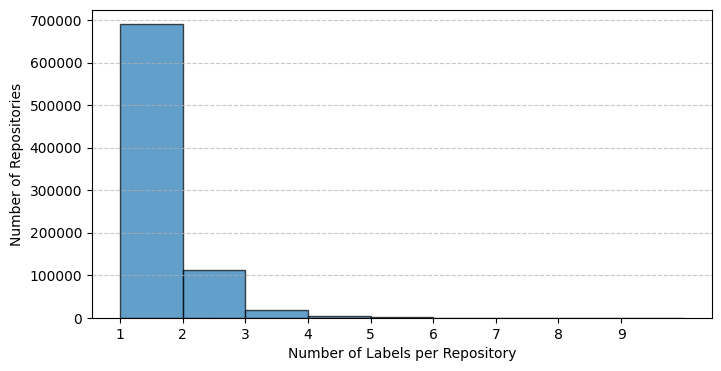}
    \caption{Distribution of the number of labels per repository. Most repositories have only one or two associated labels, showing the sparsity of topic annotations.}
    \label{fig:labels_per_repo}
\end{figure}

The number of labels per repository is also highly skewed (Figure~\ref{fig:labels_per_repo}): 83.6\% have one topic, 13.6\% have two, 2.3\% have three, 0.4\% have four, and only 0.1\% have five.
 This reflects the general sparsity of topic annotations.

\section{Methodology}
\label{sec:methodology}

\name is based on the BERT machine learning model, trained on the dataset of Section \ref{sec:dataset}, and evaluated against the RQs above.
We describe in this section the high-level building blocks of \name, leaving the technical details of its training to Section \ref{sec:experimental_setup}.

\subsection{Classification pipeline}
\label{subsec:classification_pipeline}

\name is trained on two primary inputs:
\begin{itemize}
\item \emph{File tree data}: repository name plus up to $k{=}50$ distinct file and directory names sampled uniformly from the repository’s tree; names are cleaned (trimmed, empty dropped) and deduplicated per repository, with casing preserved. Sampling is deterministic per repository and sampled names are sorted for stable artifacts. This keeps inputs within BERT’s 512-token budget and leaves room for the README when present.
\item \emph{Textual input}: The README content, if available.
\end{itemize}
We applied the BERT tokenizer’s default behavior: if the combined length of the two segments exceeds the 512-token limit, tokens are truncated from the longer segment first.

We feed these inputs into a BERT model configured for \emph{sentence-pair} classification \citep{devlin_bert_2019}.
This design leverages segment embeddings to distinguish the file tree (first segment) from the README content (second segment).
The final BERT \texttt{[CLS]} token representation encodes information across both segments.
A fully connected layer with a sigmoid activation produces probability scores for each of the 239 topic labels.
This separation of file-tree and README inputs is designed to preserve robustness when documentation is absent, directly supporting our investigation of RQ2.

To address the high class-imbalance present in the dataset, we use \emph{focal loss} \citep{Lin_focal_loss2017}, which down-weights easy examples and places greater emphasis on \emph{hard-to-classify} instances.
Formally, for a given label $y \in {0,1}$ with predicted probability $p$, the focal loss is expressed as: \begin{equation} \text{FL}(p) = -\alpha (1-p)^\gamma \text{BCE}(x,y) \end{equation} where $\text{BCE}(x,y)$ is the binary cross-entropy, $\alpha \in [0,1]$ is a weighting factor for the positive class, and $\gamma \geq 0$ controls the strength of down-weighting.
High values of $\gamma$ focus the loss more heavily on difficult or misclassified examples.
This strategy not only mitigates skewed label frequencies, but also helps stabilize gradients for rare labels, leading to better overall performance.

\subsection{Model architecture and design rationale}
\label{subsec:model_choice}

Modern transformer encoders provide a wide design space. DistilBERT \citep{sanh_distilbert_2019}, for instance, trades some accuracy for faster inference, while very long-context variants are unnecessary in our case, as both the file-tree sample and the README text comfortably fit within BERT’s 512-token window. We therefore adopt the original BERT architecture \citep{devlin_bert_2019} as a practical balance between quality and efficiency.

The choice of a sentence-pair formulation, is motivated by robustness considerations. Multi-modal learning research shows that naive early fusion (simple concatenation) allows a dominant modality to suppress weaker ones, reducing performance when one input is missing \citep{baltrusaitis_multimodal_2019,gao_asymmetry_2025}. By contrast, segment embeddings act as explicit boundaries: if the README segment is empty, the model still receives a clean representation of the file-tree input and does not conflate “missing” with “short.”

Empirical evidence supports this design. CodeBERT encodes source code and documentation as two segments separated by \texttt{[SEP]}, preserving distinct modality information while still enabling cross-attention \citep{fend_codebert_2020}. Similar paired-segment formulations consistently outperform single-sequence concatenation in natural-language inference and question-answering tasks—the same benchmarks on which BERT was originally validated \citep{devlin_bert_2019}.

Our own experiments (Section \ref{subsec:error_similarity}) confirm these findings: the sentence-pair encoding maintains high performance even when the README is absent, whereas naive concatenation leads to measurable degradation. This architectural choice directly contributes to the robustness goals defined in RQ2.

\subsection{Thresholding strategies}
\label{subsec:thresholding_concepts}

To deal with the fact that the dataset rarely contains more than three labels per repository, we introduce a \emph{quality filter} on top of the predicted probabilities, inspired by previous work \citep{dang_legion_2024}.
This filter prevents low-confidence classes from being assigned, improving precision and ensuring that only the most relevant topics are predicted.

To implement this filtering mechanism, we explore two thresholding strategies: a \emph{single global threshold}, which uniformly removes low-confidence predictions, and a \emph{per-class threshold}, which assigns a distinct confidence threshold to each label based on its frequency and distribution, in details:

\begin{itemize}

\item \textit{Single threshold:} as done in prior work \citep{dang_legion_2024}, a \textit{single threshold} selects the top-$k$ predicted labels for each repository, discarding those with confidence below a predefined common value $\tau$.

\item \textit{Per-class thresholds:} A single global threshold may not be optimal for all labels, as different topics exhibit varying levels of confidence distribution and frequency in the dataset.
To take this into account, per-class thresholds are distinct values $\tau_c$ for each topic $c$.
Predictions for every class are initially produced, yielding a predicted probability $P(c)$.
Then all classes are ranked in descending order by $P(c)$, and the selection of the top k prediction is denoted as $O_{\text{origin}}^{k}$.
For each class $c$ in $O_{\text{origin}}^{k}$ we apply the corresponding $\tau_c$, retaining only classes whose predicted probability is greater or equal than the respective thresholds.
The final set of topics $O$ for a given repository is hence:
\begin{equation}
    O = \{ c \in O_{\text{origin}}^{k} \| P(c) \geq  \tau_c \}
\end{equation}

Each per-class threshold $\tau_c$ is tuned independently using a validation set, optimizing for a balance between precision and recall to maximize the overall F1-score and class coverage \citep{Narasimhan_class_cov_2021,Ding_coverage_2023}.
This strategy allows us to account for the varying confidence distributions across different labels, yielding finer control over multi-label predictions.
The precise tuning methodology is detailed in Section \ref{sec:experimental_setup}.
\end{itemize}

\section{Experimental setup}
\label{sec:experimental_setup}

We now detail the experimental setup of training \name, covering: train/validation/test partitioning, thresholding, comparison baselines, and evaluation metrics.

\subsection{Dataset partition}
\label{subsec:data_splitting}

We randomly partition the final dataset from Section \ref{sec:dataset} of \datasetRepos repositories into three training/validation/testing datasets using the following ratios: 81\%, 9\%, and 10\%.

\subsection{Training configuration}
\label{subsec:training_config}

Unless otherwise stated, we fine-tune \texttt{BERT-base-uncased} with a multi-label classification head using focal loss. We use the Hugging Face tokenizer for \texttt{bert-base-uncased}. Inputs are sentence pairs: Segment A concatenates the repository name with up to 50 sampled file and directory names, Segment B is the README text when available.

We use \texttt{max\_length} equal to 512 with pairwise truncation set to \emph{longest first}. When the pair exceeds the budget, tokens are removed from the longest segment, which preserves the shorter segment intact.

We train with AdamW and a linear learning rate schedule using the default learning rate of \num{5e-5}. Mixed precision is enabled. We set per device train and evaluation batch size equal to 2 and use gradient accumulation with 4 steps, yielding an effective batch size of 8 per device. We train for 3 epochs. Other AdamW parameters follow the library defaults.

We use focal loss with $\alpha=0.5$ and $\gamma=2.0$ applied to the logits produced by the classification head. This down weights easy examples and emphasizes hard and rare labels.

We evaluate at each epoch end and save a checkpoint at epoch end, keeping the best checkpoint by micro \emph{F1@5}. We reload the best model at training end.

During validation we compute precision, recall, and F1 at $k \in \{1,3,5\}$ with micro averaging, as well as threshold based scores used later for threshold tuning. The selection metric for early model choice is micro \emph{F1@5}.

Imbalance is handled by the loss, we do not use data re sampling in the default setting. We also experiment with the resampling loss from LEGION in a separate variant and with a weighted BCE variant; both use the same data and evaluation pipeline as the focal loss model.

We fix the random seed for the tokenizer and data loader in our training scripts and report exact configurations in the replication package, including model name, all \texttt{TrainingArguments}, and per model hyperparameters.

\subsection{Evaluation metrics}
\label{subsec:evaluation_metrics}

We evaluate repository classification using standard multi-label metrics, including precision@k, recall@k, and F1@k for $k \in [1,5]$, and report \emph{micro-averaged} scores to capture predictive accuracy.
We include top-$k$ metrics because most repositories have few labels (83.6\% with one, 13.6\% with two; see Section \ref{par:stats}). For deployment alignment, we also consider threshold-only evaluation without fixing $k$ when discussing thresholding strategies.

We also measure class coverage \citep{Narasimhan_class_cov_2021,Ding_coverage_2023}, which assesses how well the model distributes its predictions across the ground-truth categories, preventing bias toward frequent labels. We define it as:
\begin{equation}
\text{ClassCoverage} = \frac{\left| \{ c \mid \sum_{i=1}^{N} \hat{y}_{i,c} > 0 \} \right|}{\left| \{ c \mid \sum_{i=1}^{N} y_{i,c} > 0 \} \right|}
\end{equation}
where \( y_{i,c} \) and \( \hat{y}_{i,c} \) represent the true and predicted labels for class \( c \) in instance \( i \), and \( N \) is the total number of instances.

\subsection{Threshold tuning}
\label{subsec:thresholding}

We tune the various thresholds for the approaches described in Section \ref{subsec:thresholding_concepts} as follows:
\begin{itemize}
    \item \textit{Global threshold:} to tune the global threshold we perform a grid search over $\tau \in [0.0, 1.0]$ on the validation set, selecting the threshold that maximizes F1@5, following the approach of \citet{dang_legion_2024}.
    \item \textit{Per-class thresholds}: we optimize per-class thresholds using coordinate descent \citep{wright_coordinate_descent_15}, an iterative method that updates one variable at a time while keeping others fixed. The goal is to balance the micro-averaged F1-score with class coverage, ensuring a broader distribution of predicted classes.

The process starts by initializing all thresholds to a default value. Then, for a fixed number of iterations, we cycle through each class \( c \), evaluating candidate thresholds \( \tau_c \in [0.0, 1.0] \). At each step, \( \tau_c \) is temporarily updated while others remain unchanged, and model predictions are re-evaluated to compute micro F1-score and class coverage. The objective function is:
\begin{equation}
    L= \text{F1}_{\text{micro}} + \lambda_{\text{class}} \cdot \text{ClassCoverage}
\end{equation}
where $\lambda_{\text{class}}$ is a weighting factor promoting diverse class predictions. The optimal threshold for each class is the candidate maximizing this function.

Once all classes have been updated for a full cycle, the algorithm evaluates whether the improvement in the composite score exceeds a predefined tolerance.
If the improvement is negligible, the optimization stops early; otherwise, another iteration is performed.
\end{itemize}

\subsection{Baselines}
\label{subsec:validation_baselines}

We compare our approach against both internal and external baselines.
Internally, we validate our methodology by re-implementing LEGION \citep{dang_legion_2024} from its original replication package\footnote{\url{https://github.com/AI4Code-HUST/LEGION}} and training it on our dataset.
This guarantees an equitable comparison between LEGION and \name: performance differences will be attributable to methodological advancements only, rather than dataset discrepancies.

As external baselines, we train our model on the benchmark introduced by LEGION to evaluate all state-of-the-art repositories classifiers, up to that point in the literature, namely:
\begin{itemize}
\item \textbf{LEGION} \citep{dang_legion_2024} itself, the best performing repository classifier in the literature (before \name);
\item \textbf{ZestXML} \citep{widyasari_topic_2023}, a classifier optimized for long-tail label distributions;
\item \textbf{TF-IDF+LR} \citep{izadi_topic_2021} (or just \textbf{LR} in the following, for conciseness), that combines file tree and textual features using a logistic regression classifier.
\end{itemize}
The benchmark introduced by LEGION to compare these classifiers, which we use as external baselines, consists of \num{15262} repositories crawled from GitHub, annotated with 665 unique labels.
This dataset contains, for each repository: repository name, README file (in general a well-curated one), and associated topics.
Unlike our dataset, LEGION does not include file tree information, and its topics also cover programming languages—both factors that \emph{disadvantage} \name\ relative to other baselines.

\begin{figure}
    \centering
    \includegraphics[width=0.99\linewidth]{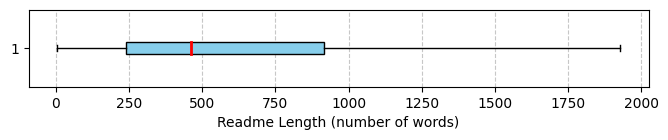}
    \caption{Distribution of README lengths (in words) in the LEGION dataset. Compared to our dataset (median 75), LEGION READMEs are substantially longer (median 463 words), indicating higher documentation quality.}
    \label{fig:readme_legion}
\end{figure}

Figure~\ref{fig:readme_legion} shows the distribution of README lengths in the LEGION dataset. Compared to our dataset (median of 75 words), LEGION’s READMEs are considerably more detailed, with a median of 463 words, an interquartile range from 238 to 914, and no missing READMEs. This confirms that LEGION repositories tend to be well-documented, which can benefit models that rely heavily on textual features.

By evaluating on both standard benchmark and real-world repositories (with, among other variations, frequently missing README files), we assess the robustness of \name across very diverse conditions.

\section{Results}
\label{sec:results}

We breakdown the presentation of our experimental results by research question (see Section \ref{sec:intro}).

\subsection{Evaluation on large-scale repository collections (RQ1)}
\label{sub:res_rq1}

We evaluate \name on our \emph{domain-only} dataset, which excludes programming language topics and contains repositories that often lack comprehensive READMEs, thereby providing a realistic and challenging setting for repository classification.
Our goal is to assess whether the approach of \name, designed to handle sparse project descriptions, can achieve state-of-the-art performance in identifying high-value topics.

Table \ref{tab:rq1summary} reports precision@k, recall@k, and F1@k for $k \in \{1,5\}$ under three thresholding methods (cf. Section \ref{subsec:thresholding_concepts}): \emph{no threshold}, \emph{single global threshold}, and \emph{per-class thresholds}.

\begin{itemize}
\item \textbf{No threshold} simply ranks the top-$k$ predictions without filtering, capturing the highest recall (53.6\% at $k=1$, 82.1\% at $k=5$), but at the cost of many spurious predictions that degrade F1 (down to 31.8\% at $k=5$).

\item \textbf{Single global threshold} imposes a single confidence cutoff, boosting precision and yielding the best overall F1 scores at both $k=1$ (59.3\%) and $k=5$ (60.8\%).
Even a modest threshold effectively discards low-confidence labels while retaining enough moderate-confidence predictions to maintain a solid recall.

\item \textbf{Per-class thresholds} slightly trail the single global threshold in F1 (58.1\% vs.\ 60.8\% at $k=5$), but achieve broader \emph{class coverage}.
Across the 239 topics in our dataset, the single global threshold leaves 15.1\% of topics without predictions, while no-threshold and per-class thresholding reduce this to 6.5\% and 6.9\%, respectively.
The per-class thresholds approach balances filtering low-confidence predictions while \emph{cutting topic loss by more than half}, ensuring better coverage of rare topics without excessive noise.
\end{itemize}

\begin{table}
\caption{Comparison of \name variants vs LEGION on our domain-only dataset (without programming languages). All values are percentages, and improvements are reported as relative percentages.}
\label{tab:rq1summary}
\centering
\begin{tabular}{@{ }l@{ }c@{ }c@{ }c|c@{ }c@{ }c@{ }}
\toprule
\multirow{2}{*}{\textbf{Model}} & \multicolumn{3}{c|}{\textbf{$k=1$}} & \multicolumn{3}{c}{\textbf{$k=5$}} \\
\cmidrule(lr){2-7}
& \textbf{P} & \textbf{R} & \textbf{F1} & \textbf{P} & \textbf{R} & \textbf{F1} \\
\midrule
\name, no threshold & 64.2 & \textbf{53.6} & 58.4 & 19.7 & \textbf{82.1} & 31.8 \\
\name, per-class threshold & 70.1 & 50.9 & 59.0 & 57.6 & 58.6 & 58.1 \\
\name, single threshold & 68.7 & 52.1 & \textbf{59.3} & 62.0 & 59.7 & \textbf{60.8} \\
\midrule
LEGION \cite{dang_legion_2024} & \textbf{77.1} & 42.5 & 54.8 & \textbf{72.0} & 44.2 & 54.8 \\
\midrule
\name improvement (\%) & -10.9 & +22.6 & +8.2 & -13.9 & +35.1 & +10.9 \\
\bottomrule
\end{tabular}
\end{table}

\subsubsection{Comparison with LEGION}
To further evaluate \name we also conducted a head-to-head comparison against LEGION, the current state of the art in repository classification, under the same domain-only conditions of the previous experiments.
As shown in Table \ref{tab:rq1summary}, LEGION achieves the highest precision at both $k=1$ (77.1\%) and $k=5$ (72.0\%), but its recall is substantially lower (42.5\% and 44.2\%, respectively).
Our single global threshold variant, by contrast, maintains a stronger balance, translating into the top F1 scores at $k=1$ (59.3\%) and $k=5$ (60.8\%).
Performing better than LEGION under identical training conditions positions \name as a state-of-the-art approach for large-scale repository classification in diverse environments where one cannot rely on high-signal information like README files.

\vspace{0.5em}

\noindent
\textit{Answer to RQ1:} on domain-only labels at scale, \name achieves F1@5 = 60.8\%, exceeding LEGION's 54.8\% by +6.0 points (Table \ref{tab:rq1summary}).

\subsection{Evaluation on curated repository datasets (external validity)}
\label{subsec:comparison_sota}

Although our primary objective lies in classifying large-scale collections of repositories with sparse documentation and domain-only labels, we also evaluate our pipeline on the more conventional curated dataset used in prior repository classification work \citep{dang_legion_2024,izadi_topic_2021,widyasari_topic_2023}.
The most recent version of this benchmark is from the LEGION paper \citep{dang_legion_2024}.

This dataset diverges from our intended use case in three key aspects: (1) each repository includes a \emph{README}, (2) \emph{programming languages} are included in the target labels, and (3) \emph{file tree data} are missing.
Even if these aspects put \name at a disadvantage, testing it on an established curated benchmark helps establish its generality.

\subsubsection{Validating our LEGION replication}
As a prerequisite to comparing \name with other methods, we first re-implemented LEGION, starting from its replication package and training it on the benchmark dataset.
Table \ref{tab:legion_comparison} shows the original scores from the LEGION paper compared to our replicated results.
Although our implementation shows a slight drop in recall but a marginally higher overall $F1$ on some ranks, these numbers remain close to or exceed the originally reported performances, validating our reimplementation.
These results also further validate our earlier comparison on our domain-only dataset: observed gains of \name over LEGION cannot be attributed to a flawed reimplementation.

\begin{table}
    \caption{Comparison of original LEGION results vs. our own replication of them for the following metrics: precision (P), recall (R), and F1 score at $k=1,3,5$. All values are percentages. Improvements are reported as relative percentages.}
    \label{tab:legion_comparison}
    \centering
    \begin{tabular}{@{}l@{ } c@{ } c@{ } c | c@{ } c@{ } c | c@{ } c@{ } c@{ }}
    \toprule
    & \multicolumn{3}{c|}{\textbf{$k=1$}} & \multicolumn{3}{c|}{\textbf{$k=3$}} & \multicolumn{3}{c}{\textbf{$k=5$}} \\
    \cmidrule(lr){2-4} \cmidrule(lr){5-7} \cmidrule(lr){8-10}
    \textbf{Model}
    & P & R & F1 & P & R & F1 & P & R & F1 \\
    \midrule
    LEGION (orig.) & 74.4 & 29.3 & 42.1 & 61.6 & 45.1 & 52.1 & 60.0 & 46.7 & 52.5 \\
    LEGION (repl.) & 76.3 & 28.6 & 41.6 & 68.0 & 43.5 & 53.1 & 67.2 & 44.6 & 53.6 \\
    \midrule
    Improvement (\%) & +2.6 & -2.4 & -1.2 & +14 & -3.5 & +1.9 & +12 & -4.5 & +2.1 \\
    \bottomrule
    \end{tabular}
\end{table}

\begin{table}
    \caption{Comparison of state-of-the-art approaches for repository classification on the LEGION dataset. Reported values are F1 scores at $k=1,3,5$. All values are percentages, and improvements are reported as relative percentages.}
    \label{tab:sota_all}
    \centering
    \begin{tabular}{l c c c}
    \toprule
    \textbf{Model} & F1@1 & F1@3 & F1@5 \\
    \midrule
    ZestXML \cite{widyasari_topic_2023} & 37.9 & 46.5 & 41.6 \\
    LR \cite{izadi_topic_2021}      & 38.8 & 50.7 & 50.0 \\
    LEGION \cite{dang_legion_2024}  & \textbf{42.1} & \textbf{52.1} & \textbf{52.5} \\
    \name  & 40.8 & 51.5 & 52.1 \\
    \midrule
    Improvement (\%) & -3.1 & -1.2 & -0.8 \\
    \bottomrule
    \end{tabular}
\end{table}

\subsubsection{\name vs.\ established baselines}
Table \ref{tab:sota_all} compares \name with other representative baselines, including ZestXML \citep{widyasari_topic_2023}, LR \citep{izadi_topic_2021}, and LEGION \citep{dang_legion_2024} (original results from the paper, not our replication).
Although LEGION achieves the highest $F1$ on this more README-rich dataset (e.g., F1@1 = 42.1\%, F1@5 = 52.5\%), \name remains highly competitive, trailing by only 0.8\% at $F1@5$ in relative terms.
While \name is not the top performer under these more polished conditions than the reality of large-scale collections, it still surpasses other strong baselines by a considerable margin.

\smallskip
In sum, \name demonstrates near state-of-the-art performance on a benchmark where: every repository has a README, file tree data are not available, and programming languages are part of the labels.
The fact that \name closely matches LEGION in such a setting, remarkably different from our target scenario, underscores its adaptability.
Meanwhile, on \textbf{a large-scale and ``messy'' corpus, \name clearly surpasses LEGION}, achieving higher recall and, consequently, higher $F1$ scores.

\subsection{Enhancing multi-label classification for README-scarce repositories (RQ2)}
\label{sub:res_rq2}

In real-world repositories, documentation quality varies widely: many projects include detailed READMEs, while others provide none at all. A practical classifier must therefore remain effective under both conditions. We refer to this ability as \emph{robustness}, meaning that the model maintains strong predictive performance when README content is missing, without requiring retraining or specialized variants. Such robustness ensures that a single model can adapt seamlessly to repositories of different completeness, which is essential for large-scale collections like Software Heritage.

To verify this, we employ a \emph{sentence-pair BERT model} (SP-BERT), which keeps file tree and README inputs as separate segments, and compare it with a standard \emph{BERT} model that concatenates them into a single sequence. This design addresses two real-world challenges: (1) 34\% of repositories lack a README; (2) 75\% of present READMEs are shorter than 200 words. Below, we compare the two approaches in three scenarios, establishing that \textbf{SP-BERT consistently outperforms its single-sequence counterpart} when README data are scarce.

\subsubsection{Full test set results}
Table \ref{tab:rq2_summary} compares F1 scores at $k\in\{1,3,5\}$ for SP-BERT and standard BERT across the three thresholding strategies.
The single global threshold yields the highest overall F1 in both models, but SP-BERT provides an additional gain at every $k$ (e.g., $F1@5$: 59.7\% $\rightarrow$ 60.8\%).
This suggests that maintaining file tree and textual signals as distinct segments helps the classifier learn more robust features, even when READMEs are scarce.

\begin{table*}
\setlength{\tabcolsep}{3pt}
\caption{Sentence pair BERT (SP BERT) versus traditional BERT under three thresholding strategies. Values are F1 scores at $k\in\{1,3,5\}$, as percentages. The last row shows the relative improvement of SP BERT over BERT.}
\label{tab:rq2_summary}
\centering
\begin{tabular}{lccc|ccc|ccc}
\toprule
& \multicolumn{3}{c|}{\textbf{No Threshold}}
& \multicolumn{3}{c|}{\textbf{Single Threshold}}
& \multicolumn{3}{c}{\textbf{Per-Class Threshold}} \\
\cmidrule(lr){2-4}\cmidrule(lr){5-7}\cmidrule(lr){8-10}
\textbf{Model} & $F1@1$ & $F1@3$ & $F1@5$
               & $F1@1$ & $F1@3$ & $F1@5$
               & $F1@1$ & $F1@3$ & $F1@5$ \\
\midrule
BERT    & 57.2 & 42.3 & 31.4  & 58.1 & 59.7 & 59.7  & 57.7 & 57.4 & 56.9 \\
SP BERT & 58.4 & 42.9 & 31.8  & 59.3 & 60.8 & 60.8  & 59.0 & 58.6 & 58.1 \\
\midrule
Improvement (\%) & +2.1 & +1.4 & +1.3  & +2.1 & +1.8 & +1.8  & +2.3 & +2.1 & +2.1 \\
\bottomrule
\end{tabular}
\end{table*}

\subsubsection{Forced README removal}
To measure the impact of keeping the two inputs separate, we evaluate a scenario where we forcibly remove READMEs from all repositories in the test set.
Table \ref{tab:rq2_remove} reports F1@1\ldots5 for both models.
Without any textual data at inference, SP-BERT still outperforms standard BERT by a clear margin (e.g., single global threshold at $k=5$: 53.9\% $\rightarrow$ 55.2\%).
This indicates that training the model with separate file tree and README segments encourages more robust learning from file tree data, maintaining strong performance when README is missing at test time.

When comparing \name's performance using the entire dataset to its performance after removing the README at \emph{test time}, the results show a decrease of only 5.7\% (from F1@1 = 59.3\% to 53.6\%) and 5.6\% (from F1@5 = 60.8\% to 55.2\%). In relative terms, this represents a drop of approximately 9.6\% for F1@1 and 9.2\% for F1@5.
A similar experiment by \citet{izadi_topic_2021} investigates the effect of removing README and wiki data, but at \emph{training time}.
Even with that tailored training regime, they reported an F1@5 drop from 47.0\% to 37.4\%—a relative loss of more than 20\%.
In contrast, our method operates \emph{without} re-training and sees only a relative 9.2\% decline.

\begin{table*}
\setlength{\tabcolsep}{3pt}
    \caption{Forced README removal: F1 Scores at $k \in \{1,3,5\}$ for sentence-pair BERT (SP-BERT) and traditional BERT. All values are percentages and the last row shows the relative improvement of SP-BERT over BERT.}
    \label{tab:rq2_remove}
    \centering
    \begin{tabular}{lccc|ccc|ccc}
    \toprule
    & \multicolumn{3}{c|}{\textbf{No Threshold}}
    & \multicolumn{3}{c|}{\textbf{Single Threshold}}
    & \multicolumn{3}{c}{\textbf{Per-Class Threshold}} \\
    \cmidrule(lr){2-4} \cmidrule(lr){5-7} \cmidrule(lr){8-10}
    \textbf{Model} & $F1@1$ & $F1@3$ & $F1@5$
                   & $F1@1$ & $F1@3$ & $F1@5$
                   & $F1@1$ & $F1@3$ & $F1@5$ \\
    \midrule
    BERT
                   & 51.1 & 38.3 & 28.8
                   & 52.3 & 53.9 & 53.9
                   & 47.6 & 47.9 & 47.5 \\
    SP-BERT
                   & 52.4 & 38.9 & 29.2
                   & 53.6 & 55.2 & 55.2
                   & 53.1 & 52.6 & 52.1 \\
    \midrule
    Improvement (\%)
                   & +2.5 & +1.6 & +1.4
                   & +2.5 & +2.4 & +2.4
                   & +11.6 & +9.8 & +9.7 \\
    \bottomrule
    \end{tabular}
\end{table*}

\subsubsection{Real-world README-less repositories}
Finally, we measure performance on the \emph{actual} subset of repositories that lack a README in our dataset (34\%).
This real-world scenario (Table \ref{tab:rq2_real_noreadme}) confirms the trend: SP-BERT retains higher F1 scores across all $k$, compared to standard BERT.
Notably, under single global thresholding at $k=5$, SP-BERT reaches F1 = 52.1\%; standard BERT remains at 50.7\%.
Even on real-world README-less repositories, training with a dedicated second segment confers better robustness.

\vspace{0.5em}
\noindent
\textit{Answer to RQ2:} DRAGON maintains robustness without retraining under README removal (\emph{F1@5 drop} $\approx$ 9.2\% relative). Across all tested conditions and thresholding policies, sentence-pair BERT (file-tree | README) achieves higher F1 than single-sequence BERT; the margin is $\ge$2 points at $k{=}5$ and is more pronounced when README is missing, underscoring the benefit of treating file-tree signals as a separate input segment for README-scarce settings.

\begin{table*}
\setlength{\tabcolsep}{3pt}
    \caption{Real-world README-less repositories: F1 Scores at $k \in \{1,3,5\}$ for sentence-pair BERT (SP-BERT) and traditional BERT. All values are percentages and the last row shows the relative improvement of SP-BERT over BERT.}
    \centering
    \label{tab:rq2_real_noreadme}
    \begin{tabular}{lccc|ccc|ccc}
    \toprule
    & \multicolumn{3}{c|}{\textbf{No Threshold}}
    & \multicolumn{3}{c|}{\textbf{Single Threshold}}
    & \multicolumn{3}{c}{\textbf{Per-Class Threshold}} \\
    \cmidrule(lr){2-4} \cmidrule(lr){5-7} \cmidrule(lr){8-10}
    \textbf{Model} & $F1@1$ & $F1@3$ & $F1@5$
                   & $F1@1$ & $F1@3$ & $F1@5$
                   & $F1@1$ & $F1@3$ & $F1@5$ \\
    \midrule
    BERT
                   & 47.6 & 37.8 & 28.9
                   & 48.4 & 50.7 & 50.7
                   & 47.6 & 47.9 & 47.5 \\
    SP-BERT
                   & 48.9 & 38.4 & 29.2
                   & 49.9 & 52.2 & 52.1
                   & 49.3 & 49.5 & 49.0 \\
    \midrule
    Improvement (\%)
                   & +2.7 & +1.6 & +1.0
                   & +3.1 & +3.0 & +2.8
                   & +3.6 & +3.3 & +3.2 \\
    \bottomrule
    \end{tabular}
\end{table*}

\subsection{Thresholding trade-offs and evaluation policy (RQ3)}
\label{subsec:rq3_results}

We evaluate three thresholding strategies---\emph{no threshold}, \emph{single global threshold}, and \emph{per-class thresholds}---across all experimental settings introduced in Sections~\ref{sub:res_rq1} and~\ref{sub:res_rq2}.
This analysis quantifies how thresholding affects model performance under different input and model conditions, including the absence of README files.

\subsubsection{Domain-only dataset}
Table \ref{tab:rq1summary} summarizes the main results for our large-scale \emph{domain-only} dataset, which excludes programming languages and includes many repositories with sparse documentation.

Here, thresholding strongly impacts the balance between precision and recall.
The \emph{no-threshold} ranking achieves the highest recall (82.1\% at $k=5$) but yields the lowest F1 (31.8\%), confirming that unrestricted ranking introduces substantial noise.
Applying a \emph{single global threshold} improves overall balance, reaching the best F1 scores at both $k=1$ (59.3\%) and $k=5$ (60.8\%), as low-confidence predictions are effectively filtered.
\emph{Per-class thresholds} slightly trail in F1 (58.1\% at $k=5$) but increase topic diversity: the share of unpredicted topics drops from 15.1\% (single threshold) to 6.9\%.
Thus, threshold selection directly controls the precision--coverage trade-off.

\subsubsection{SP-BERT vs.\ BERT}
Table \ref{tab:rq2_summary} compares the effect of thresholding when using \emph{SP-BERT} versus a conventional \emph{BERT}.
For the baseline BERT model, thresholding shows the same pattern observed with SP-BERT.
Without any threshold, performance remains low ($F1@5 = 31.4\%$), as unfiltered rankings introduce many incorrect labels.
Applying a \emph{single global threshold} improves performance, reaching $F1@5 = 59.7\%$, while \emph{per-class thresholds} yield a slightly lower score ($F1@5 = 56.9\%$).

\subsubsection{Impact of README availability}
We also examine how thresholding behaves when README content is missing, both in a controlled removal setting and on the real subset of README-less repositories.
Across both conditions, absolute F1 scores decrease but the relative ordering of thresholding strategies remains unchanged.
The \emph{no-threshold} variant continues to yield the lowest F1, while introducing a \emph{single global threshold} again provides the best overall performance.
At $k=5$, F1 under the single-threshold policy declines moderately—from 60.8\% with full inputs to 55.2\% after README removal, and to 52.1\% on repositories that lack README files entirely—whereas \emph{per-class thresholds} remain slightly lower.

\vspace{0.5em}
\noindent
\textit{Answer to RQ3:} a single global threshold maximizes F1 on this dataset; per-class thresholds trade a small F1 loss for substantially higher class coverage.

\section{Error and Semantic Similarity Analysis}
\label{subsec:error_similarity}

We now study where errors concentrate and whether misclassifications are semantically meaningful. This section connects four pieces: distribution of errors across labels, their aggregate contribution, the role of frequency, and the semantic structure of confusions. Together these answer RQ4.

\subsection{Distribution and nature of errors}

Unless otherwise noted, we report results as recall@5 under the single threshold policy introduced in Section \ref{subsec:thresholding_concepts}, because here we localize where errors arise by counting missed ground truth labels per class. This makes it the most appropriate metric for identifying which classes are systematically underpredicted. Later in this section, we also translate recall values into absolute miss counts by multiplying with class support.

Because the test set, as the original dataset, is strongly imbalanced in the number of ground truth topics per repository, error counts should be interpreted with that distribution in mind. In particular, \num{69 050} repositories have a single topic (83.6\%), \num{11 207} have two (13.6\%), \num{1 858} have three (2.3\%), \num{340} have four (0.4\%), and only \num{76} have five (0.1\%).

\begin{figure}
    \centering
    \includegraphics[width=0.78\linewidth]{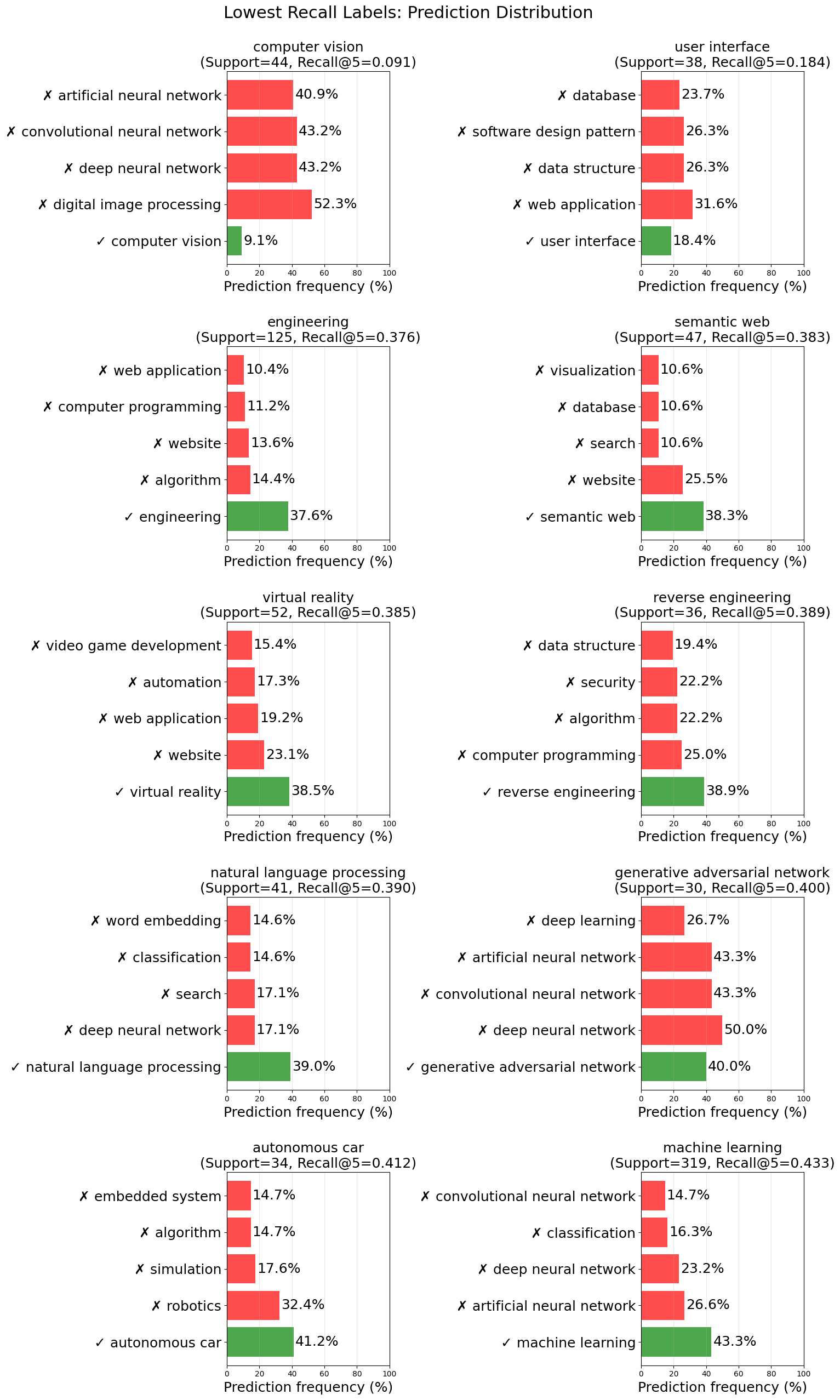}
    \caption{Prediction distribution analysis for lowest performing labels. For each problematic label, the chart shows the percentage of correct predictions and the most frequent incorrect predictions.}
    \label{fig:lowest_recall_dist}
\end{figure}

\subsubsection{Lowest recall classes}
We first report the labels with the lowest absolute recall@5, restricting to classes with at least ten test instances. Table \ref{tab:lowest_recall} lists the top-30. Most of these classes are infrequent, but some with meaningful support, such as \emph{computer vision} and \emph{engineering}, also show low recall. The distribution plots in Figure \ref{fig:lowest_recall_dist} reveal which alternative labels the model tends to predict in these cases. For example:
\begin{itemize}
    \item \emph{computer vision} (support 44, recall 0.091) is most often confused with \emph{digital image processing} (52.3\%), \emph{deep neural network} (43.2\%), and \emph{artificial neural network} (40.9\%).
    \item \emph{engineering} (support 125, recall 0.376) is replaced by broad categories such as \emph{algorithm} (14.4\%), \emph{website} (13.6\%), and \emph{web application} (10.4\%).
    \item Other low recall labels such as \emph{user interface} are confused with \emph{web application} (31.6\%), \emph{data structure} (26.3\%), and \emph{software design pattern} (26.3\%).
    \item \emph{natural language processing} is substituted by \emph{deep neural network} (17.1\%), \emph{search} (17.1\%), and \emph{classification} (14.6\%).
    \item \emph{autonomous car} tends to be replaced with \emph{robotics} (32.4\%), \emph{simulation} (17.6\%), or \emph{algorithm} (14.7\%).
    \item Even high profile research topics such as \emph{generative adversarial network} are frequently confused with \emph{deep learning} (50\%) and \emph{artificial neural network} (43.3\%).
\end{itemize}

\begin{table}[ht]
\centering
\small
\caption{Top-30 topics with the lowest recall@5 with support at least ten. The label shown in bold appears in both tables.}
\label{tab:lowest_recall}
\begin{tabular}{lrrr}
\toprule
\textbf{Label} & \textbf{Support} & \textbf{Recall@5} & \textbf{Missed} \\
\midrule
social network & 13 & 0.000 & 13 \\
artificial intelligence & 20 & 0.000 & 20 \\
computational biology & 13 & 0.000 & 13 \\
computer vision & 44 & 0.091 & 40 \\
graph database & 11 & 0.091 & 10 \\
functional programming & 10 & 0.100 & 9 \\
electronic trading platform & 26 & 0.115 & 23 \\
computer science & 24 & 0.125 & 21 \\
knowledge graph & 15 & 0.133 & 13 \\
user interface & 38 & 0.184 & 31 \\
telecommunications network & 16 & 0.188 & 13 \\
geographic information system & 18 & 0.278 & 13 \\
long short term memory & 12 & 0.333 & 8 \\
text mining & 12 & 0.333 & 8 \\
generative model & 24 & 0.333 & 16 \\
collaborative filtering & 20 & 0.350 & 13 \\
augmented reality & 16 & 0.375 & 10 \\
engineering & 125 & 0.376 & 78 \\
semantic web & 47 & 0.383 & 29 \\
virtual reality & 52 & 0.385 & 32 \\
reverse engineering & 36 & 0.389 & 22 \\
natural language processing & 41 & 0.390 & 25 \\
operating system & 10 & 0.400 & 6 \\
generative adversarial network & 30 & 0.400 & 18 \\
image editing & 17 & 0.412 & 10 \\
autonomous car & 34 & 0.412 & 20 \\
autonomous driving & 14 & 0.429 & 8 \\
\textbf{machine learning} & \textbf{319} & \textbf{0.433} & \textbf{181} \\
face detection & 20 & 0.450 & 11 \\
hyperparameter optimization & 51 & 0.451 & 28 \\
\bottomrule
\end{tabular}
\end{table}

\subsubsection{Main error contributors}
The previous analysis focused on labels with the lowest recall, i.e., those the model struggles to predict correctly. However, such labels may be rare and therefore have little impact on the overall performance. We now shift focus to identify which labels \emph{contribute most to the total number of errors} across the test set.
To measure this, we estimate for each class $c$ the number of ground-truth instances that were missed by the model, computed as:

\[
\text{Missed}(c) = \text{support}(c) \times \bigl(1 - \text{recall@5}(c)\bigr).
\]
This quantity represents the absolute count of missed labels for each class and highlights labels whose residual errors dominate the aggregate error budget, even if their individual recall is moderate.
Table \ref{tab:error_contribution} shows the top contributors. Broad categories such as \emph{web application}, \emph{database}, \emph{website}, and \emph{server} dominate the error budget. \emph{Machine learning} appears in both the lowest recall table and the highest error contributors, reflecting its combination of difficulty and prevalence. Distribution plots in Figure \ref{fig:lowest_recall_dist} confirm that \emph{machine learning} is often substituted with \emph{artificial neural network} (26.6\%), \emph{deep neural network} (23.2\%), or \emph{classification} (16.3\%).

\begin{table}[ht]
\centering
\small
\caption{Top-30 error contributing topics ranked by number of misses with support at least ten. The label shown in bold appears in both tables.}
\label{tab:error_contribution}
\begin{tabular}{lrrr}
\toprule
\textbf{Label} & \textbf{Support} & \textbf{Recall@5} & \textbf{Missed} \\
\midrule
web application & 1763 & 0.778 & 391 \\
database & 2222 & 0.830 & 377 \\
website & 4144 & 0.911 & 367 \\
server & 1231 & 0.755 & 301 \\
data & 982 & 0.697 & 298 \\
automation & 1318 & 0.775 & 297 \\
algorithm & 2916 & 0.906 & 273 \\
search & 758 & 0.646 & 268 \\
computer programming & 693 & 0.620 & 263 \\
security & 1267 & 0.800 & 254 \\
visualization & 1259 & 0.799 & 253 \\
education & 558 & 0.548 & 252 \\
front end & 1365 & 0.818 & 248 \\
back end & 873 & 0.729 & 237 \\
software testing & 1167 & 0.805 & 227 \\
web application security & 407 & 0.459 & 220 \\
mathematics & 1007 & 0.799 & 202 \\
statistics & 797 & 0.749 & 200 \\
game & 5085 & 0.961 & 199 \\
animation & 1013 & 0.808 & 195 \\
client & 571 & 0.664 & 192 \\
design & 594 & 0.687 & 186 \\
mathematical optimization & 613 & 0.701 & 183 \\
analytics & 529 & 0.658 & 181 \\
\textbf{machine learning} & \textbf{319} & \textbf{0.433} & \textbf{181} \\
digital image processing & 1141 & 0.842 & 180 \\
image & 886 & 0.798 & 179 \\
real time computing & 426 & 0.580 & 179 \\
websocket & 1041 & 0.829 & 178 \\
classification & 846 & 0.790 & 178 \\
\bottomrule
\end{tabular}
\end{table}

\subsubsection{Concentration and frequency effects}
Errors are not uniformly spread. The top-30 error contributing labels account for \num{7 139} of \num{17 695} misses, that is 40.3\% of all errors, and the top-10 alone account for 17.5\%. In parallel, label frequency is a first order driver of performance: across the 239 labels we observe a strong positive association between log support and recall@5 with Pearson correlation of 0.859. Thus, frequency improves recall on average yet also magnifies the absolute number of misses whenever any residual error remains. Rare classes are systematically brittle. Among labels with average support near 6, the mean recall@1 is 0.034 and the mean recall@5 is 0.133, with 33 labels at recall@1 equal to zero and 23 labels at recall@5 equal to zero.
\begin{figure}
    \centering
    \includegraphics[width=1\linewidth]{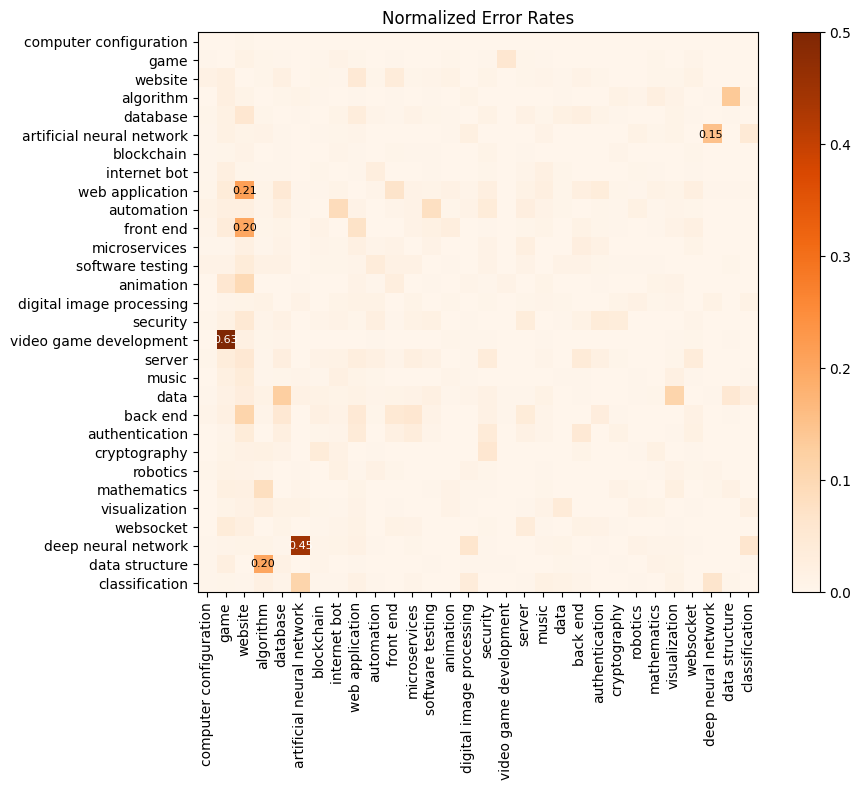}
    \caption{Each cell shows the proportion of true instances of a label predicted as another, after removing the diagonal.
Rows are normalized \emph{before} diagonal removal, so values represent actual error rates per class rather than relative shares among errors, highlighting how much of each class leaks into others.}
    \label{fig:confusion_matrix}
\end{figure}

\subsection{Semantic structure of confusions}
We next examine whether misclassifications are semantically close to the target labels. This connects the error distributions above with the semantic organization of the label space.

\subsubsection{Observed confusions}
Figure \ref{fig:confusion_matrix} is a confusion matrix restricted to the thirty most frequent classes. The figure shows that many wrong predictions are semantically close to the true label. For instance, repositories annotated as \emph{web application} are often predicted as \emph{website} or \emph{server}, and \emph{database} is often replaced by \emph{data} or \emph{analytics}. For low recall classes, distribution plots in Figure \ref{fig:lowest_recall_dist} highlight that the model confuses \emph{natural language processing} with \emph{deep neural network} and \emph{classification}, \emph{autonomous car} with \emph{robotics}, and \emph{computer vision} with \emph{digital image processing}. These substitutions indicate that errors concentrate within topical neighborhoods rather than being unrelated.

\subsubsection{Critical groups of labels}
Taken together, the previous results delineate where effort should focus next. The findings consistently point to three groups:
\begin{enumerate}
    \item Rare classes with low recall such as \emph{artificial intelligence} or \emph{computational biology}.
    \item Common classes with moderate deficits that generate many absolute errors such as \emph{web application}, \emph{database}, and \emph{machine learning}.
    \item Overlapping classes prone to mutual confusion such as \emph{web application} versus \emph{website}, or \emph{computer vision} versus \emph{digital image processing}.
\end{enumerate}
This typology motivates the next analysis, where we examine whether these confusions reflect meaningful semantic proximity rather than arbitrary mistakes.

\subsubsection{Semantic similarity of labels}
To examine whether these confusions reflect semantic proximity, we embed all label names with a SentenceTransformer and compute cosine similarity between pairs. The similarity matrix reveals clusters of related terms. High scoring examples include \emph{debugger} and \emph{debugging} with 0.91, \emph{deep learning} and \emph{deep neural network} with 0.87, \emph{data visualization} and \emph{visualization} with 0.86, and \emph{autonomous car} and \emph{autonomous driving} with 0.85. These are candidates for merging or aliasing prior to training and evaluation.

\subsubsection{Similarity and confusion}
For the thirty most frequent labels we correlate pairwise similarity with empirical confusion counts. We find a positive and statistically significant association (Pearson $r=0.34$, Spearman $\rho=0.39$, $p < 10^{-25}$).
In other words, labels that are closer in embedding space are more likely to be substituted in predictions. Examples include \emph{video game development} predicted as \emph{game} in 487 cases with similarity 0.48, and \emph{deep neural network} predicted as \emph{artificial neural network} in 213 cases with similarity 0.72. Other high similarity confusions include \emph{front end} versus \emph{back end} with 0.76 and \emph{data} versus \emph{data structure} with 0.70.

\begin{figure}
    \centering
    \includegraphics[width=0.75\linewidth]{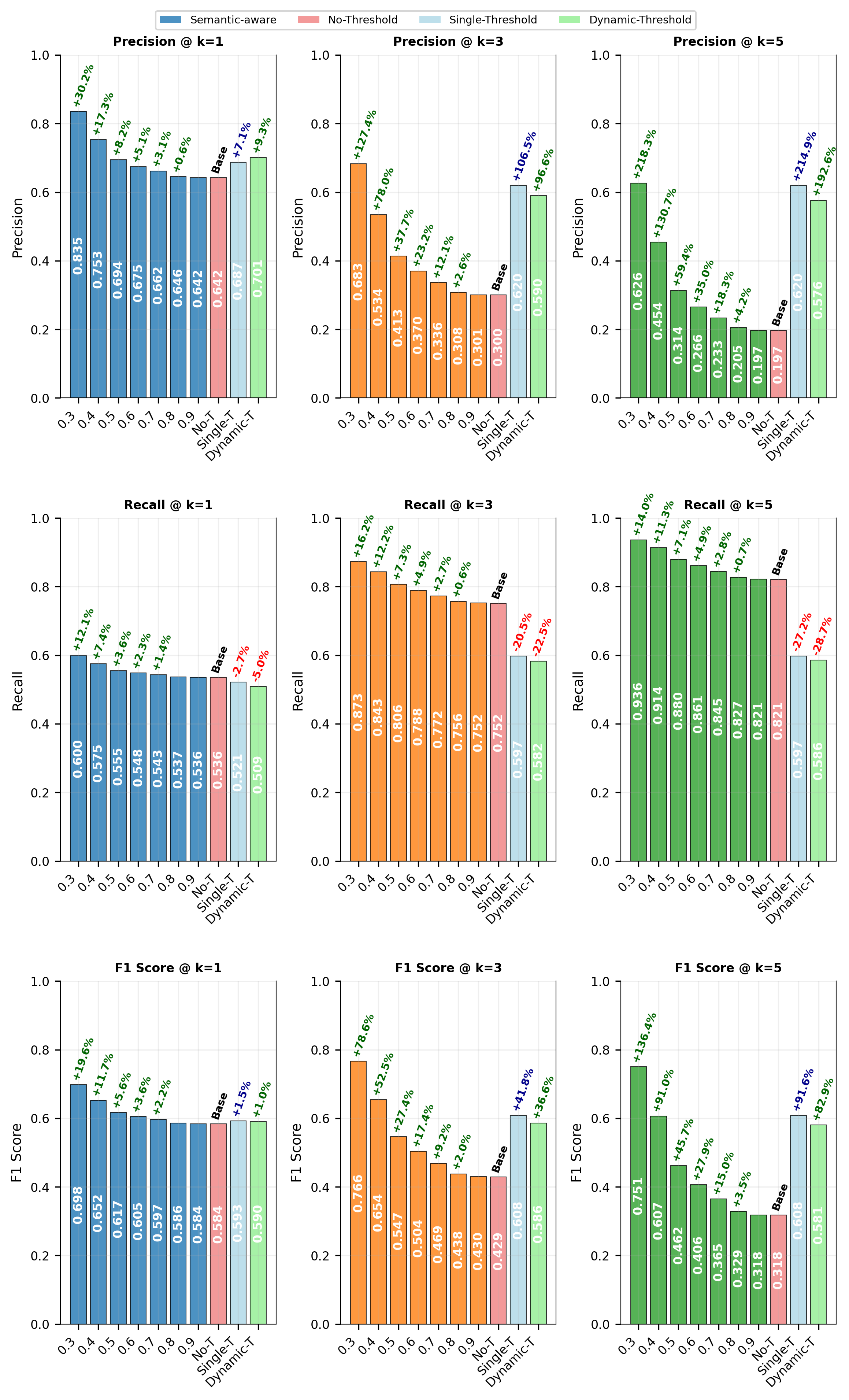}
    \caption{Comparison of evaluation methods across top-k settings. Rows correspond to metrics and columns to top-k values. Bars show scores for semantic-aware thresholds alongside Baseline (No Threshold), Single-Threshold, and Dynamic-Threshold methods. Numeric labels report absolute scores; annotations above bars report percent improvement relative to the Baseline.}
    \label{fig:semantic-evaluation}
\end{figure}

\subsubsection{Synonym aware evaluation}

Having established that confusions correlate with similarity, we now examine whether the most error-prone classes occur in semantically dense regions of the label space. To quantify this, we measure the average embedding similarity among the top error-contributing classes and compare it to the global label space.
Labels that dominate the error budget tend to occur in dense semantic regions. The mean similarity among the top error contributing classes is 0.3354, compared to 0.2025 across all label pairs. Between 14\% and 16.4\% of all apparent errors can be reinterpreted as synonym confusions if pairs with similarity above 0.5 are treated as equivalent, which indicates that many errors are near misses in meaning.

Building on this observation, we perform a relaxed evaluation that explicitly accounts for such semantic proximity by accepting synonyms above a similarity threshold as correct. This allows us to estimate how many of the observed “errors” are actually semantically valid alternatives rather than true misclassifications. To help interpret the effect of different thresholds more clearly, Figure~\ref{fig:semantic-evaluation} visualizes the changes in precision, recall, and F1 across top-k settings (k = 1, 3, 5). Gains persist across different similarity thresholds: from 0.5 to 0.7. At 0.5, precision@5 increases from 0.197 to 0.314 and recall@5 from 0.821 to 0.880 with \num{48 245} corrections. At 0.6, precision@5 rises to 0.266 and recall@5 to 0.861 with \num{28 438} corrections. Even at 0.7, precision@5 improves to 0.233 and recall@5 to 0.845 with \num{14 883} corrections. Moreover, false positives are often close to the ground truth in embedding space. At rank one, the mean similarity of false positives to their true labels is 0.345 with 14.6\% above 0.5, and the same pattern holds for ranks three and five with 16.2\% and 14.6\%.

\vspace{0.5em}
\noindent
\textit{Answer to RQ4:} frequent labels achieve higher recall on average, yet because they appear often and lie in dense semantic neighborhoods, their residual mistakes account for most of the absolute error budget and are frequently near synonyms of the ground truth. Between 14\% and 16.4\% at rank five can be reclassified as acceptable under synonym aware evaluation, and similarity correlates with confusion, indicating that many residual errors are semantically close rather than arbitrary.

\section{Discussion}
\label{sec:class_discussion}

\subsection{On the need of a ``messy'', real-world dataset for large-scale repository classification}

The performances of automated classifiers depend heavily on the dataset(s) they are tested on.
When initially faced with the need of classifying repositories from very large real-world collections, we quickly realized that previous approaches in the state-of-the-art had not been benchmarked on comparable datasets, resulting in unmet expectations.

To address this, we assembled a realistic dataset, corresponding to the target problem.

The first design choice behind the dataset is to exclude programming languages as target labels because, on the one hand, they are trivial to detect with existing heuristics and tools and, on the other, their presence would inflate performance metrics without yielding deeper domain insights about input repositories.

The second design choice is to avoid relying on all user-provided labels, like GitHub topics, because they can be inconsistent, redundant, or both.
Instead, we adopted the curated taxonomy of GitRanking \citep{sas_gitranking_2023}, which offers a structured set of meaningful domain-specific topics, mitigating the noise associated with crowdsourced annotations.

Third, most existing approaches rely heavily on README content, in contrast with the real-world of large repository collections where README files are often either absent (34\% of the repositories in our dataset lack a README, up to 42\% of the entire SWH archive) or very short.
By incorporating these README-scarce repositories in our dataset, we encourage the development of classifiers that perform well in README-scare contexts.

Finally, we also wanted a \emph{both large-scale and real-world dataset}, in order to be able to scale to scenarios like SWH, where 96\% of the archived 350 million repositories do not have any topic.

Testing \name and baselines on different benchmarks shed light on the applicability of repository classifiers in different contexts.

LEGION \citep{dang_legion_2024} achives slightly higher scores on a polished dataset where every repository has a README and topics include programming languages.
But on a real-world ``messy'' benchmark, with scarce README information and only predicting high-value topics, \name outperforms LEGION by 8.2\% at F1@1 and 10.9\% at F1@5 in relative terms, ultimately achieving F1@1 = 59.3\% and F1@5 = 60.8\%. These results underscore \name's robustness in the real-world conditions of large repository collections, where other methods have never been tested.

Significant work went in the creation of our dataset, which is an important byproduct of the development of \name; \textbf{we strongly encourage future work on repository classification to benchmark against it}.

\subsection{Sentence-pair BERT is key for handling README scarcity}

Our evaluation in Section \ref{sub:res_rq2} shows that sentence-pair BERT improves classification performance when README content is missing.
Across all test scenarios (full dataset, forced README removal, real-world README-less repositories) \name outperforms standard BERT concatenation.
The effect is particularly evident when READMEs are absent, where sentence-pair modeling effectively leverages file-tree data, whereas naive concatenation in traditional BERT leads to a significant drop in predictive power.

A key advantage of \name is its ability to maintain classification performance without requiring specialized re-training.
In contrast, \citet{izadi_topic_2021} observed that removing README and wiki content during training led to an F1@5 drop from 47.0\% to 37.4\%---a relative decline of 20\%.
\name, even \emph{without re-training}, experiences only a 9.2\% F1@5 relative loss when README files are removed at test time, nearly halving the gap in a much more challenging scenario.
This demonstrates how effectively \name exploits file-tree data, making it more robust for real-world collections where READMEs are scarce.

These results suggest that the file-tree structure of a repository provides a latent signal sufficient for meaningful classification, even without textual descriptions.
Thanks to its sentence-pair architecture, \name captures these signals more effectively than prior approaches.
This aligns with findings from related work in different contexts \citep{rokon_repo2vec_2021,zhang_higit_2019}, where separating structural and textual information also improved software representation learning.

\subsection{Thresholding strategies: precision vs. coverage trade-offs}

The analysis of thresholding strategies in Section \ref{sub:res_rq1} reveals that single-threshold filtering consistently yields the best overall F1 scores.
This finding aligns with prior work from \citet{dang_legion_2024}, showing that filtering out low-confidence predictions helps prevent spurious topic assignments.

However, our per-class thresholding approach provides a crucial advantage: broader class coverage.
While it slightly underperforms in raw F1 score (58.1\% vs. 60.8\% at F1@5), it ensures that more distinct topics appear in model predictions. The impact is significant: the number of unpredicted topics \emph{shrinks by more than half} (15.1\% vs 6.9\%).
This suggests a precision vs. class coverage trade-off: stricter filtering improves accuracy at the cost of reducing diversity in the assigned labels. Our approach allows prospective users to reason on this tradeoff, which depends on the target application: if precision is paramount (e.g., automated tagging systems), a single-threshold approach is preferable; if maximizing class coverage is critical (e.g., exploratory search engines), per-class thresholds may be more appropriate. Finally, by pointing out the issue, we encourage future work to improve on this target.

\subsection{What errors and similarity tell us about evaluation and utility}

 The error analysis shows two concentrations of mistakes. First, rare labels have low recall even with careful sampling and focal loss. Second, broad and frequent labels such as \textit{web application} or \textit{database} accumulate many absolute errors even when recall is acceptable.

The similarity analysis clarifies that many substitutions are semantically close to the ground truth. Confusions are more likely between labels that are close in embedding space, and the classes that dominate the error budget lie in dense semantic regions. When we accept close synonyms above a fixed similarity threshold, between \num{14}\% and \num{16.4}\% of apparent errors at rank five are reclassified as acceptable matches. This complementary view does not replace strict metrics but helps interpret them: strict metrics favor conservative selectivity, whereas synonym aware views favor semantic coverage. For future works, a hierarchical or merged taxonomy for very close concepts could reduce artificial confusion and improve interpretability while keeping strict metrics as the primary score.

\subsection{Why Top-5 suggestions are useful with sparse ground truth}

Many repositories in our data have very few annotated topics
(83.6\% have 1 topic, 13.6\% have 2, 2.3\% have 3, 0.4\% have 4 and only 0.1\% have 5).
This sparsity means that the ground truth often captures only a narrow slice of a project.
In practice, a model can propose additional labels that are reasonable for the same repository even if they are not listed in the annotations.

Our similarity analysis shows that a large share of apparent errors are semantically close to the target labels.
Hence, a short list of Top five suggestions is valuable in real workflows such as tagging, curation, or search enrichment.
Even when not all five match the annotated ground truth, several suggestions are near synonyms or closely related concepts and therefore useful to users.

\subsection{Threats to validity}
\label{sec:threats}

While \name is an effective large-scale classification pipeline, some factors may impact the generalizability, fairness, and reproducibility of our results.

\subsubsection{Internal validity}

A primary concern lies in the quality of the ground-truth labels.
Our dataset is derived from user-assigned GitHub topics mapped through the GitRanking taxonomy.
These annotations may be incomplete, repositories can lack valid topics or be noisy, with incorrect or overly generic labels.
Such inconsistencies could bias reported metrics by overstating or understating true performance.
Although GitRanking reduces this risk through curation and normalization, it cannot eliminate the original label noise entirely.

Another potential threat is the reliance on a single train–validation–test split.
As is customary for large-scale transformer-based models, we do not perform multiple resampling runs.
This may cause moderate variance in reported scores depending on the initial data partitioning.
Future work could explore repeated stratified sampling or cross-validation to assess the sensitivity of results.

Finally, while focal loss mitigates class imbalance, some residual bias toward frequent topics may persist, affecting per-class recall estimates.

To ensure methodological transparency, we used standard evaluation metrics (scikit-learn), the official Hugging Face BERT implementation, and verified our replication of LEGION against its original results.
All code, data, and trained models are publicly available through our replication package \citep{replication-package}.

\subsubsection{External validity}

 While our dataset is designed to reflect large-scale software archives with missing or minimal README files, models trained on it might underperform on repositories that rely heavily on curated textual documentation.

To mitigate this, we also tested our methodology on the benchmark dataset used in prior work.

\section{Conclusion}
\label{sec:conclusion}

In this article, we introduced \name, a robust and scalable multi label repository classifier that assigns high value, domain specific topics at scale. \name excludes programming languages since they are easy to infer by other means, and instead targets application and domain labels that are more useful for discovery. The method operates with light inputs that are preserved in version control, combining file and directory names with optional README text through a sentence pair BERT encoder, and addresses class imbalance with focal loss.

The approach is designed for scarce data scenarios that arise in exhaustive archives, where source code level mining is impractical at scale and READMEs are frequently absent. In Software Heritage, 42\% of repositories lack a README, yet \name remains effective without retraining, showing about a 9\% relative drop in F1@5 when README is removed at inference. On domain only labels and at large scale, \name achieves F1@5 equal to 60.8\% and improves over the best prior approach by about 11\% in relative terms. An analysis of errors shows that most mistakes are semantic near misses among closely related topics. Thresholding studies further show that a single global threshold maximizes micro F1, while per class thresholds can be used to increase topic diversity and class coverage, which is valuable for exploratory search.

A key contribution of this work is the largest dataset to date for repository topic classification, with more than 825 thousand repositories mapped to 239 GitRanking topics. We release the dataset together with code and models to support replication and reuse. Since these 825 thousand repositories represent the entire set of labeled projects among more than 200 million preserved in Software Heritage, over 99\% of repositories remain unlabeled today, which highlights the need for automated classification approaches such as ours.

Few limitations remain. Some labels are rare or overlapping which complicates strict evaluation. The tokenizer and cleaning choices favor English, which can affect multilingual projects. Future work includes taxonomy consolidation to reduce overlap, multilingual modeling, strategies for the long tail, confidence calibration, and human in the loop workflows. Taken together, these directions can further strengthen \name as a practical building block for navigating and mining very large collections of open source software.

\end{document}